\documentclass[onecolumn,12pt]{article}
\usepackage{jheppub}

\usepackage[font={small}]{caption}
\usepackage{subfigure}
\usepackage{graphicx}
\usepackage{dcolumn}
\usepackage{float}
\usepackage[normalem]{ulem}

\usepackage{mathrsfs} 

\usepackage{multicol}
\usepackage{cancel}
\usepackage{mathtools}
\usepackage{hyperref}

\usepackage{subfigure}
\usepackage{xcolor}
\def\({\left(} \def\){\right)}
\def\[{\left[} \def\]{\right]}

\newcommand{\eg}{{\it e.g.,}\ }
\newcommand{\ie}{{\it i.e.,}\ }

\newcommand{\tr}{\text{Tr}}

\newcommand{\beq}{\begin{equation}}
\newcommand{\eeq}{\end{equation}}
\newcommand{\bea}{\begin{eqnarray}}
\newcommand{\eea}{\end{eqnarray}}

\newcommand{\cre}{charged R\'enyi entropy}

\newcommand{\cres}{charged R\'enyi entropies}

\def\le{\left(}
\def\ri{\right)}
\def\del          {\partial}

\def\tr           {\mathop{\rm Tr}}

\renewcommand{\eqref}[1]{(\ref{#1})}

\begin{document}

\title{Shape Deformations of Charged R\'enyi Entropies from Holography}

\author{Stefano Baiguera$^{1}$, Lorenzo Bianchi,$^{2,3}$ Shira Chapman$^{1}$ and Dami\'an A. Galante$^4$}

\affiliation{$^1$ Department of Physics, Ben-Gurion University of the Negev, \\ David Ben Gurion Boulevard 1, Beer Sheva 84105, Israel}
\affiliation{$^2$ Universit\'a di Torino, Dipartimento di Fisica,\\ Via P. Giuria 1, I-10125 Torino, Italy}
\affiliation{$^3$ INFN. - sezione di Torino,\\ Via P. Giuria 1, I-10125 Torino, Italy}
\affiliation{$^4$ Department of Mathematics, King’s College London,\\ the Strand, London WC2R 2LS, UK}

\emailAdd{baiguera@post.bgu.ac.il}
\emailAdd{lorenzo.bianchi@unito.it}

\emailAdd{schapman@bgu.ac.il}\emailAdd{damian.galante@kcl.ac.uk}

\abstract{
\sloppy Charged and  symmetry-resolved R\'enyi entropies are entanglement measures quantifying the degree of entanglement within different charge sectors of a theory with a conserved global charge. 
We use holography to determine the  dependence of charged R\'enyi entropies on small shape deformations away from a spherical or planar entangling surface in general dimensions. This dependence is completely characterized by a single coefficient appearing in the two point function of the displacement operator associated with the R\'enyi defect. We extract this coefficient using its relation to the one point function of the stress tensor in the presence of a deformed entangling surface. This is mapped to a holographic calculation in the background of a deformed charged black hole with hyperbolic horizon. We obtain numerical solutions for different values of the chemical potential and replica number $n$ in various spacetime dimensions, as well as analytic expressions for small chemical potential near $n=1$. When the R\'enyi defect becomes supersymmetric, we demonstrate a conjectured relation between the two point function of the displacement operator and the conformal weight of the twist operator.}

\maketitle

\setcounter{tocdepth}{2}

\section{Introduction}\label{introduction}

Quantum states differ from classical ones due to the presence of entanglement. In order to characterize this entanglement, several measures were proposed, including the entanglement and R\'enyi entropies. However, in systems with a global symmetry, it turns out that these measures do not capture the intricate interplay between different charge sectors of the theory. For this reason, a new measure was proposed -- the charged R\'enyi entropy 
\begin{equation}\label{CREmain}
S_n(\mu) \equiv \frac{1}{1-n} \log \tr \left[\rho_A\frac{e^{\mu Q_A}}{n_A(\mu)}\right]^n
\end{equation}
where $\rho_A=\tr_{\bar A}\rho$ is the reduced density matrix of the state of the full system $\rho$ over a region $A$, $Q_A$ is the charge operator restricted to this region, $\mu$ is a chemical potential conjugate to the charge $Q_A$  and $n_A (\mu)\equiv \tr \left[\rho_A\, e^{\mu Q_A}\right]$ is a normalization constant.  
We assume that the (pure) state of the full system is an eigenstate of the charge operator, \ie $[\rho,Q]=0$ and therefore $[\rho_A,Q_A]=0$. This means that the reduced density matrix can be decomposed in blocks corresponding to the different charge sectors $\rho_A=\oplus p(q)\rho_A(q)$, with probabilities $p(q)$  of finding an outcome $q$ when measuring $Q_A$ and $\tr\rho_A(q)=1$.

The evaluation of charged R\'enyi entropies can be rephrased in terms of a partition function
\begin{equation}\label{intro:partfunmu}
    S_n(\mu) = \frac{1}{n-1}\left(n \log Z_1(\mu) - \log Z_n(\mu)\right), \qquad Z_n(\mu) \equiv \tr \left[\rho_A e^{ \mu Q_A}\right]^n.
\end{equation}
Then, the symmetry-resolved R\'enyi entropies 
\begin{equation}\label{symmetryresolvedRN}
    S_n(q) \equiv \frac{1}{1-n}\log \tr \rho_A(q)^n
\end{equation}
within each given charge sector $q$, are obtained by performing a Laplace transform over this partition function  \cite{Bonsignori:2019naz}\footnote{We see in this equation that $\mu \equiv i\mu_E$ is a natural redefinition. In field theory situations the parameter $\mu$ is typically imaginary, and in this case it simply produces a phase for charged fields as they go around the entangling surface in the replicated geometry often used to define R\'enyi entropies \cite{Goldstein:2017bua}. On the other hand, in holography it can be either real or imaginary. We have selected a convention consistent with the previous holographic literature \cite{Belin:2013uta}.  \label{physicalchem}}  
\begin{equation}\label{symmetryresolvedRN2}
   S_n(q) = \frac{1}{n-1}\left(n \log \mathcal{Z}_1(q) - \log \mathcal{Z}_n(q)\right) 
   \qquad \mathcal{Z}_n(q) \equiv 
   -i n \int_{-i\pi/n}^{i\pi/n} \frac{d\mu}{2\pi} e^{- q n \mu} Z_n(\mu),
\end{equation}
and the probability to be in a given charge sector is $p(q) = \mathcal{Z}_1(q)$. 
Charged R\'enyi entropies have recently gained a lot of attention within the condensed matter literature, \eg \cite{Goldstein:2017bua,Cornfeld:2018wbg,Bonsignori:2019naz,Murciano:2019wdl,Azses:2020wfx,Capizzi:2020jed,Parez:2021pgq,Calabrese:2021wvi,PhysRevA.100.022324,PhysRevLett.121.150501,Estienne:2020txv,Oblak:2021nbj}. In particular, they were proposed as a useful tool for identifying symmetry-protected topological (SPT) states \cite{Matsuura:2016qqu,Azses:2020wfx}.\footnote{SPT states are states which are connected to a product state by a local unitary transformation, but this transformation breaks the symmetry of the system.} This observation was used to detect  SPT states on the IBM quantum computer by implementing a protocol which measures the second charged R\'enyi entropy \cite{Azses:2020tdz}.

In conformal field theories in $d$ spacetime dimensions, the partition function in eq.~\eqref{intro:partfunmu} can be evaluated by performing a Euclidean path integral on a replicated geometry with twisted boundary conditions along the entangling surface. The twisted boundary conditions can be implemented by inserting a $d-2$ dimensional twist operator in the Euclidean path integral \cite{Calabrese:2004eu,Calabrese:2005zw,Hung:2014npa}. For a charged theory, the twist operator is further dressed by a magnetic flux fixed by the parameter $\mu$. 

When the entangling surface is spherical or planar, the charged R\'enyi entropies take a particularly simple form and can be evaluated analytically in various examples of free and holographic CFTs  \cite{Belin:2013uta}. More generally, such a shape of the entangling surface preserves a large subgroup of the original conformal symmetry, therefore turning our CFT to a defect CFT (dCFT) \cite{Billo:2016cpy,Bianchi:2015liz}. Since some of the symmetry has been broken, dCFTs are not as powerful as CFTs. Nevertheless, many correlation functions are completely fixed by the symmetry. For example, dCFTs are equipped with a primary operator living on the defect - the displacement operator - which, when inserted in correlation functions, displaces the defect locally. The kinematics of several correlation functions involving this operator are completely fixed by symmetry and this fact will play a crucial role in our analysis.

The goal of the present paper is to study the shape dependence of the charged R\'enyi entropy for small deformations away from a symmetric entangling surface -- spherical or planar -- in theories with a holographic dual. In dCFTs, such deformations are completely characterized in terms of a single  coefficient $C_D$, which appears in the two point function of the displacement operator  -- see eq.~\eqref{eq:two_point_function_DD} below. We will compute this coefficient for holographic theories in various spacetime dimensions $d$, as a function of the R\'enyi index $n$ and the chemical potential $\mu$. In order to achieve this goal, we employ the fact that $C_D$ also appears in the one point function of the stress tensor in the presence of a deformed defect.

The holographic setup for studying charged R\'enyi entropies with a planar or spherical entangling surface includes charged black holes with hyperbolic horizons. Indeed, this quantity was studied extensively in holography, see \eg  \cite{Belin:2013uta,Lewkowycz:2013nqa,Caputa:2013eka,Wong:2013gua,Belin:2014mva,Pastras:2014oka,Matsuura:2016qqu,Zhao:2020qmn,Weisenberger:2021eby,Zhao:2022wnp}. 
Since we are interested in small deformations of  symmetric entangling surfaces, our background will consist of a slightly deformed version of the charged hyperbolic black hole. The coefficient $C_D$ will be extracted by evaluating the one point function of the stress tensor in this deformed background.

The R\'enyi defect can be tuned such that it preserves supersymmetry (SUSY) and in this case the charged R\'enyi entropies become  supersymmetric. This particular case was extensively studied in \cite{Nishioka:2013haa,Nishioka:2014mwa,Hama:2014iea,Nishioka:2016guu,Hosseini:2019and} and an interesting conjecture was made \cite{Bianchi:2019sxz} suggesting that the coefficient $C_D$ and the conformal weight of the twist  operator $h_n$ 
are proportional, \ie
\begin{equation}
C_D^{\rm conj} (n,\mu) = d \, \Gamma \le \frac{d+1}{2} \ri \le \frac{2}{\sqrt{\pi}} \ri^{d-1} h_n (\mu) \, .
\label{eq:conj_CD}
\end{equation}
This was rigorously established in $d=4$ \cite{Bianchi:2019sxz} and, after the initial proposal of \cite{Lewkowycz:2013laa,Fiol:2015spa}, it has been extensively checked for supersymmetric Wilson lines in $d=3$ \cite{Drukker:2019bev}.\footnote{In the three-dimensional case the twist operator is a line defect, and we can extend to the supersymmetric R\'enyi entropy the results that have been derived for supersymmetric Wilson lines using only the preserved supersymmetry.} In this paper we demonstrate that this relation holds for holographic theories in all dimensions $3\leq d \leq 7$, when the gravity solution is tuned to be supersymmetric. 

Contrary to the uncharged case, we will find that the dCFT data $C_D(n,\mu)$ and $h_n(\mu)$ will not vanish for $n\to 1$. This is expected because the presence of a magnetic flux along a codimension-two surface, in the ordinary single-copy CFT, creates a non-trivial defect, which is commonly known as monodromy defect \cite{Billo:2013jda,Gaiotto:2013nva,Giombi:2021uae,Bianchi:2021snj,Dowker:2021gqj}. Therefore, as a byproduct of our analysis, we also study the shape dependence of holographic monodromy defects.

The paper is organized as follows. In \S \ref{sect-QFT_story} we review the various defect-CFT ingredients required to analyse the R\'enyi defects. In particular we evaluate the one point function of the stress tensor and current in the presence of deformed defect and relate it to $C_D$. In \S \ref{sec:holo} we present the holographic study of the charged hyperbolic black hole where we also evaluate the one point function of the stress tensor and extract $C_D$ in different dimensions and for various values of the R\'enyi index and chemical potential. We present analytic expansions along with numerical results and discuss the supersymmetric case. Finally, in \S\ref{discussion}, we summarize our findings and present some future directions. 
We relegate several technical details to the appendices.
Appendix \ref{app:example_3d} deals with the simple example of deformations of a circular entangling surface in three dimensions. 
In appendix \ref{app:on_shell_action} we compute the renormalized on-shell action for charged hyperbolic black holes in Einstein-Maxwell gravity. Details of the holographic renormalization method are given in appendix \ref{app:details_holo_ren}. In appendix \ref{app:diff_eq_higherd} we present the analytic and numerical computations for $C_D$ in higher dimensions.

\section{The QFT story} 
\label{sect-QFT_story}

Here, we introduce the field theory ingredients required for evaluating charged Rényi entropies in the presence of small shape deformations of a flat (or spherical) entangling surface.
We begin in section \ref{sect-replica_twist} by considering planar (or spherical) defects. We introduce generalized twist operators and explain their relation to \cres. We write the expectation values of the stress tensor and current in the presence of a planar defect.
In section \ref{sect-deformations_defect}, we describe how to perform small deformations of a planar (or spherical) entangling surface.
For this purpose, we introduce the displacement operator and review its relevant correlation functions. We explain how to use the displacement operator to derive the one-point functions of the stress tensor and current in the presence of a deformed entangling surface. Finally, in section \ref{sect-adapted_coord}, we transform these expectation values to a cylindrical coordinate system centered around the deformed entangling surface. This will be convenient when comparing with holography in the next section.

\subsection{Replica trick and twist operators}
\label{sect-replica_twist}

Consider a pure state described by a density matrix $\rho$ on a given time slice of a CFT in $d$ spacetime dimensions.  
We split the system to a region $A$ and its complement  $\bar A$ along an entangling surface $\Sigma$. 
We further assume that the CFT is invariant under a global U(1) symmetry associated to a conserved charge $Q$ and that the state $\rho$ is an eigenstate of the charge operator. We would like to understand how the charged and symmetry-resolved R\'enyi entropies \eqref{CREmain}-\eqref{symmetryresolvedRN} depend on the shape of the entangling surface.

In quantum field theory, charged  R\'enyi entropies with integer R\'enyi index $n$  can be related to the Euclidean path integral on a replicated geometry, consisting of $n$ copies of the field theory, glued together with appropriate boundary conditions. The effect of the chemical potential is taken into account by including a magnetic flux carried by the entangling surface, see \eg \cite{Belin:2013uta,Calabrese:2004eu}.
More specifically, we can use the relation
\beq
\tr \left[ \rho_A \, \frac{e^{\mu Q_A}}{n_A (\mu)} \right]^n = 
\frac{Z_n (\mu)}{\le Z_1 (\mu) \ri^n} \,,
\label{eq:general_replica_trick}
\eeq
where $Z_n(\mu)$ is the grand canonical  partition function, evaluated on an $n$--fold cover of flat space. The geometry includes branch cuts introduced on the region $A$ at the Euclidean time where the state lives, which we can choose to be $t_{\rm E}=0$, without loss of generality. The branch cuts connect the $i$-th copy of the field theory to the $(i \pm 1)$-th copy as $t_{\rm E} \rightarrow 0^{\mp}$.  
The effect of the chemical potential $\mu$, conjugate to the charge $Q$, is introduced using an appropriate background gauge field $B_{\mu}$. The gauge field is such that Wilson lines evaluated over any loop $\mathcal{C}$ encircling the entangling
surface are constant, \ie  $\oint_{\mathcal{C}} B = - i n \mu$.\footnote{The factor of $n$ comes from the $n$--sheeted geometry since the loop encircles the entangling surface $n$ times.}
Note that the constant $n_A (\mu) = \tr \le \rho_A e^{\mu Q_A} \ri $ in the denominator of eq.~\eqref{eq:general_replica_trick} was selected such that the expression is normalized as $n\rightarrow 1$ for any $\mu$.

The boundary conditions in the replicated geometry can be implemented by 
the insertion of a codimension--two twist operator $\tau_n$ at the entangling surface \cite{Hung:2011nu,Hung:2014npa}.
In the presence of a chemical potential, 
$\tau_n$ can be dressed with a Dirac sheet carrying the magnetic flux $-i n \mu$ to produce a generalized twist operator $\tilde{\tau}_n$ \cite{Belin:2013uta}. In fact, binding together the original twist operator and the magnetic flux is natural from a field theory perspective, since they both produce phases for charged fields as they go around the entangling surface. 
Inserting $\tilde{\tau}_n$ in correlation functions is then equivalent to evaluating them in the presence of the charged R\'enyi defect. 
The expectation value of the generalized twist operator is related to the grand canonical partition function as follows 
\beq
\langle \tilde{\tau}_n (\mu) [A] \rangle = \frac{Z_n (\mu)}{(Z_1 (0))^n} \, ,
\eeq
where of course, the normalization is such that the result is normalized in the absence of a defect, \ie when $n=1$ and $\mu=0$. In order to simplify the notation, we will, from now on, omit the explicit dependence of the generalized twist operators on the subregion $A$ and only keep track of their dependence on the chemical potential.
We can express the \cres ~\eqref{CREmain} in terms of the generalized twist operator as
\beq
S_n(\mu)=\frac{1}{1-n} (\log \langle\tilde \tau_n (\mu)\rangle-n\log \langle\tilde \tau_1(\mu)\rangle).
\eeq

To begin with, let us consider a planar entangling surface $\Sigma$.\footnote{Planar and spherical defects are related by a conformal transformation. For definiteness, we will work with a planar entangling surface. However, a similar discussion applies with minor modifications for the spherical case. The coefficient $C_D$ which we will extract is relevant for both cases. An example with a spherical entangling surface is discussed in appendix \ref{app:example_3d}.}
In this case, our codimension-two defect is conformal and the relevant correlation functions are discussed in \cite{Billo:2016cpy,Bianchi:2015liz}.
Let us denote the directions orthogonal to the defect by $x^a = \lbrace x^1, x^2 \rbrace,$ while the parallel coordinates will be denoted $y^i,$ with $i \in \lbrace 1, 2 , \dots , d-2 \rbrace$; 
we collectively denote them as $z^{\mu}=(x^a, y^i)$. We can place the defect at $x^a=0$, without loss of generality. 
In the presence of the planar defect, the one-point function of the energy-momentum tensor is fixed by conformal invariance up to a single constant as follows
\beq
\langle T_{ij} (z) \rangle_{n,\mu} = - \frac{h_n (\mu)}{2 \pi n} \frac{\delta_{ij}}{|x|^d} \, , \qquad
\langle T_{ab} (z) \rangle_{n,\mu} = \frac{h_n (\mu)}{2 \pi n} \frac{1}{|x|^d}
\le (d-1) \delta_{ab} - d \, \frac{x_a x_b}{x^2}  \ri \, ,
\label{eq:one_pt_function_stress_tensor}
\eeq
where $\langle \cdots \rangle_{n,\mu} = \langle \cdots \, \tilde{\tau}_n (\mu) \rangle $  denotes correlation functions evaluated in the presence of the generalized twist operator $\tilde{\tau}_n (\mu) .$
The quantity $h_n(\mu)$ is sometimes referred to as the dimension of the twist operator since it provides a generalized notion of conformal weight measured via the insertion of the stress tensor at a small perpendicular distance $x^a$  
from the defect. Assuming the validity of the Averaged Null Energy Condition it has been shown that $h_n(\mu)$ is positive in a unitary dCFT \cite{Jensen:2018rxu}. 
In equation \eqref{eq:one_pt_function_stress_tensor}, $T_{\mu \nu}$ refers to the stress tensor in a single copy of the CFT instead of the full replicated geometry. Therefore, the one-point function contains an additional factor of $n^{-1}$ compared to other references like eq.~(2.20) in \cite{Belin:2013uta}. 
Similarly, 
the one-point function of the global current in the presence of a planar defect takes the form  \cite{Belin:2013uta,Billo:2016cpy}
\beq
\langle J_a (z) \rangle_{n,\mu} = \frac{i \, a_n (\mu)}{2 \pi n} \, \frac{\epsilon_{ab}x^b}{|x|^d} \, , \qquad
\langle J_i (z) \rangle_{n,\mu} =0 \, .
\label{eq:original_1pt_function_current}
\eeq 
where $a_n(\mu)$ is the magnetic response of the current to the flux carried by the generalized twist operator. Here again, the factor of $n^{-1}$ appears since we use $J_\mu$ to denote the current in a single copy of the geometry.
The Levi-Civita symbol in the above correlator originates from the parity-odd magnetic flux carried by the twist operator. 
The parallel components instead vanish, since nothing  invariant under parallel translations can be constructed with the required indices. 

The coefficients $h_n (\mu)$ and $a_n (\mu)$ carry information about the defect.
Of course, for $n=1$ and $\mu=0$ the defect is not present and all the one-point functions vanish, \ie $h_1(0)=a_1(0)=0$. 
More generally, the expansion around $n=1$ is controlled by the modular Hamiltonian, which corresponds, for our configuration, to the insertion of an integrated stress tensor in the undeformed theory. This leads to a direct relation between the first-order expansion of $h_n(0)$ around $n=1$ and the stress tensor two-point function. Analogously, expansions around $\mu=0$ are controlled by insertions of the current $J_{\mu}$.
To see this explicitly, let us consider the two-point functions of the stress tensor and of the conserved current in an ordinary CFT \cite{Osborn:1993cr,Erdmenger:1996yc}
\beq
\langle T_{\mu\nu} (x) T_{\rho\sigma} (0) \rangle = \frac{C_T}{ x^{2d}} \, \mathcal{I}_{\mu\nu,\rho\sigma} (x) \, , \qquad \langle J_{\mu} (x) J_{\nu} (0) \rangle = \frac{C_V}{ x^{2(d-1)}} \, I_{\mu\nu} (x) \, ,
\label{eq:definition_CTCV}
\eeq
where 
\beq
\mathcal{I}_{\mu\nu, \rho\sigma} \equiv \frac{1}{2} \le I_{\mu\rho} I_{\nu\sigma} + I_{\mu\sigma} I_{\nu\rho} \ri - \frac{1}{d} \delta_{\mu\nu} \delta_{\rho\sigma} \, ,\qquad
I_{\mu\nu}(x) \equiv \delta_{\mu\nu} - 2 \frac{x_{\mu} x_{\nu}}{|x|^2} \, .
\label{eq:def_Istructure}
\eeq
and where
$C_T$ and $C_V$ are the corresponding central charges.
Using the modular Hamiltonian as outlined above, one finds \cite{Hung:2011nu,Hung:2014npa,Perlmutter:2013gua}
\beq
\del_n h_n(0)|_{n=1} = 2 \pi^{\frac{d}{2}+1} \frac{\Gamma \le d/2 \ri}{\Gamma (d+2)} \, C_T \, .
\label{eq:property_hn_around_n1}
\eeq
More generally, different correlation functions of the stress tensor and current govern the different orders of the Taylor expansion of $h_n(\mu)$ and $a_n(\mu)$ around $n=1$ and $\mu = 0$, see section 2.3 of  \cite{Belin:2013uta}.
In the case of the holographic CFT studied in section \ref{sec:holo} we present explicit expressions for $h_n(\mu)$ and $a_n(\mu)$, see eqs.~\eqref{eq:weight_twist_operator_with_chemical_potential} and \eqref{eq:an_from_holography} below.

\subsection{Small deformations of a flat defect}
\label{sect-deformations_defect}

The insertion of a twist operator breaks translational invariance  in the directions orthogonal to $\Sigma$, leading to the appearance of a contact term in the Ward identity corresponding to the conservation of the energy-momentum tensor, \ie\!\!\footnote{The notation $T^{\mu \nu}_{\rm tot}$ refers to the stress tensor in the full replicated theory $(\mathrm{CFT})^n ,$ which differs from the one inserted in a single copy by a factor of $n.$}
\beq\label{eq:Warddisp}
\del_{\mu} T^{\mu a}_{\rm tot} (x,y) = \delta_{\Sigma} (x) D^a (y) \, ,
\eeq
where $D^a$ is the displacement operator.
This is a local defect primary operator which implements small deformations in the shape of the defect.
Specifically, denoting with $X^{\mu} = (0, y^i)$ the location of the undeformed generalized twist operator, one can introduce an infinitesimal deformation $\delta X^{\mu}$ and the action of the displacement operator is defined through the identity
\beq
\langle D^a \cdots \rangle_{n, \mu} = n^{a}_{\mu} \frac{\delta}{\delta X_{\mu}} \langle \cdots \rangle_{n, \mu} \, ,
\eeq
where $n^{\mu}_a$ are the unit normal vectors orthogonal to the defect.

One-point functions of the displacement operator vanish for a flat (or spherical) defect.
The first non-trivial correlator is the two-point function, which reads
\beq
\langle D_a (y) D_b (y')  \rangle_{n, \mu} = \delta_{ab} \, \frac{C_D}{(y-y')^{2(d-1)}} \, .
\label{eq:two_point_function_DD}
\eeq
Since the normalization of the displacement operator is fixed by the Ward identity \eqref{eq:Warddisp}, the coefficient $C_D(n,\mu)$ is a physical piece of defect CFT data.\footnote{The displacement two-point function determines also the coefficients of a term in the boundary Weyl anomaly \cite{Herzog:2017kkj,Herzog:2017xha}.} Notice that $C_D$ is a norm and therefore, in a unitary dCFT, it is positive. In the following, when possible, we leave implicit the dependence of $C_D$ on both $n$ and $\mu$.

Let us consider the response of a defect to a perpendicular displacement of the form
\beq\label{deformation}
\delta X^{\mu} = \delta^{\mu}_a f^a \, ,
\eeq
where $f^a (y)$ is the profile of the deformation.
The physical response of the system is measured by the variation of the
partition function, according to 
\beq 
 \delta \log Z_n (\mu) = \frac{1}{2} \int_{\Sigma} dw \int_{\Sigma} dw' \, 
f^a (w) f^b (w') \langle D_a (w) D_b (w') \rangle_{n,\mu} + \mathcal{O} (f^4) \, ,
\label{varpart}
\eeq
where we discarded the first order term in the perturbation because the one-point function of the displacement operator vanishes for a flat (or spherical) defect.
The variation of the charged R\'enyi entropy can then be expressed as
\begin{equation}
    \delta S_n(\mu) = \frac{1}{n-1}\left(n \, \delta \log Z_1(\mu) - \delta \log Z_n(\mu)\right)   \, .
    \label{eq:variation_Sn}
\end{equation}
It then becomes clear that the second variation of the R\'enyi entropy under any shape deformation is completely fixed in terms of the coefficient $C_D$.
Indeed, using eq.~\eqref{eq:two_point_function_DD}, we obtain that the response of the charged R\'enyi entropy to the small deformation of a flat entangling surface reads
\beq
 \delta S_n(\mu)  = \frac{n \, C_D(1,\mu) - C_D (n,\mu)}{2(n-1)}  
 \int_{\Sigma} dw \int_{\Sigma} dw' \, 
\frac{f^a (w) f_a (w')}{(w-w')^{2(d-1)}}  + \mathcal{O} (f^4) \, .
\label{eq:variation_Sn_in_terms_of_CD}
\eeq
In particular, the variation is finite at $n=1,$ independently of the corresponding chemical potential. 
We provide in appendix \ref{app:example_3d} a simple application of this procedure to determine the variation of the partition function $\delta \log Z_n (\mu)$ in a three-dimensional case, where we deform a circle into various shapes.
The leading corrections in the chemical potential of the uncharged entanglement entropy ($n=1$) for a spherical entangling surface have been recently studied in \cite{Bueno:2022jbl}.

While it is possible to obtain $C_D$ from second-order perturbation theory applied to the R\'enyi entropy \eqref{eq:variation_Sn} using eqs.~\eqref{eq:two_point_function_DD}-\eqref{varpart}, it is more convenient to consider other observables such as the one-point functions of the currents.
Indeed, they allow to extract $C_D$ by working at first order in perturbation theory, according to the identities \cite{Billo:2016cpy}
\begin{equation}
\langle {T}_{\mu\nu}(z) \rangle_{n,\mu,f\Sigma}= \langle T_{\mu\nu}(z) \rangle_{n,\mu} - \int d^{d-2} w \langle D_a(w)T_{\mu\nu}(z) \rangle_{n,\mu} \, f^a(w)+\mathcal{O}\left(f^2\right) \, ,
\label{eq:one_point_function_Tdeformed}
\end{equation}
\beq
\langle J_{\mu} (z) \rangle_{n,\mu,f\Sigma} =
\langle J_{\mu} (z) \rangle_{n,\mu} - \int d^{d-2} w \, 
\langle D_a (w) J_{\mu} (z) \rangle_{n,\mu} \, f^a (w) + \mathcal{O} (f^2) \, . 
\label{eq:relation_one_point_functions_currents_shape_deformation}
\eeq
The subscript $f \Sigma$ refers to the background with a deformed entangling surface.

Let us start by focusing on the first order variation of current \eqref{eq:relation_one_point_functions_currents_shape_deformation}. 
Using the technology developed in \cite{Billo:2016cpy}, we determine the two-point functions of the current with the displacement operator: 
\beq
\begin{aligned}
& \langle D_a (w) J_i (z) \rangle_{n,\mu} = \frac{i a_n (\mu)}{2 n} \pi^{-\frac{d}{2}} \frac{\Gamma(d)}{\Gamma (d/2)} \, \frac{2 w_i \epsilon_{ab} x^b}{(x^2+w^2)^d} \, , & \\
& \langle D_a (w) J_b (z) \rangle_{n,\mu} = \frac{i a_n (\mu)}{2 n} \pi^{-\frac{d}{2}} \frac{\Gamma(d)}{\Gamma (d/2)} \, \frac{1}{(x^2+w^2)^{d-1}} \le \epsilon_{ab} - \, \frac{2 x_b \epsilon_{ac} x^c}{x^2+w^2}  \ri \, ,  &
\end{aligned}
\label{eq:two_point_functions_jD}
\eeq
where $a_n (\mu)$ is the coefficient introduced in eq.~\eqref{eq:original_1pt_function_current}.
In the above expressions, we have set $y^i=0$ inside $z=(x^a, y^i).$ 
With those in hand, one could try to perform the  integration in eq.~\eqref{eq:relation_one_point_functions_currents_shape_deformation}. While it cannot be performed in the general case, the singular terms in the short distance expansion around $x =0$ can be extracted explicitly.
The method is to perform the limit $|x|\rightarrow 0$ in the weak sense, \ie as an integration against a test function and rephrase the outcome as an expansion over distributions with support at $w=0$, specifically delta functions and their derivatives. The technical details of the expansion can be found in appendix A of \cite{Bianchi:2016xvf}. Here we only report those identities (often referred to as \emph{kernel formulas}) needed for our purposes: 
\beq
\begin{aligned}
& \frac{1}{(x^2+w^2)^{d-1}} = \frac{\pi^{\frac{d-2}{2}} \Gamma \le \frac{d}{2} \ri}{\Gamma (d-1)} 
\le  \frac{\delta^{d-2} (w)}{|x|^d} + \frac{\del^2 \delta^{d-2} (w)}{2(d-2) |x|^{d-2}} \ri + \dots ,&  \\
& \frac{1}{(x^2+w^2)^{d}} = \frac{\pi^{\frac{d-2}{2}} \Gamma \le \frac{d}{2} \ri}{2 \Gamma (d)} 
\le  \frac{d \, \delta^{d-2} (w)}{|x|^{d+2}} + \frac{\del^2 \delta^{d-2} (w)}{2 |x|^{d}} \ri + \dots , &  \\
& \frac{w^i}{(x^2+w^2)^{d}} = - \frac{\pi^{\frac{d-2}{2}} \Gamma \le \frac{d}{2} \ri}{2 \Gamma (d)} 
\le  \frac{\del^i \delta^{d-2} (w)}{|x|^d} + \frac{\del^i \del^2 \delta^{d-2} (w)}{2(d-2) |x|^{d-2}} \ri + \dots,  & 
\end{aligned}
\label{eq:kernel_formula}
\eeq
where the ellipsis stand for less singular terms in the radial distance from the entangling surface. 
Substituting those into eqs.~\eqref{eq:relation_one_point_functions_currents_shape_deformation}-\eqref{eq:two_point_functions_jD} and using integration by parts, one obtains the leading singularity in the one point function of the current.

The strategy to compute the first-order variation of the stress-tensor in eq.~\eqref{eq:one_point_function_Tdeformed} is similar. 
One starts from the two-point functions $\langle T D \rangle$ between the energy-momentum tensor and the displacement operator, which can be derived using general results in dCFTs \cite{Billo:2016cpy}.
Such two-point functions depend explicitly on the coefficients $C_D$, cf.~eq.~\eqref{eq:two_point_function_DD} and $h_n$, cf.~eq.~\eqref{eq:one_pt_function_stress_tensor}. 
One then extracts their leading singular behaviour by means of the kernel formulas.
This procedure is in fact unmodified compared to the case without a conserved charge studied in \cite{Bianchi:2016xvf} (cf.~eqs.~(2.8)-(2.9) and (2.12)-(2.13)). 
The only difference is the addition of the $\mu$ dependence of the coefficient $C_D$. We will therefore not repeat this part here and we refer the reader to \cite{Bianchi:2016xvf} for those expressions.

\subsection{Currents in adapted coordinates}
\label{sect-adapted_coord}

In preparation for the holographic computation of section \ref{sec:holo}, we move to cylindrical coordinates adapted to the shape of the deformed defect.
The idea is to send 
a congruence of geodesics orthogonal to the entangling surface $\Sigma$.   
Let us start with a planar defect. In this case we will denote the coordinates as $(\rho, \tau, y^i),$ where $\rho = |x^a|$ is the radial distance to the defect and $\tau$ is the angle in the plane perpendicular to the defect. Explicitly, the coordinates $\rho$ and $\tau$ are related to the Cartesian coordinates $x^a$ as 
\beq\label{cylindcoord1}
x^a = \rho \left( \cos [ \tau/R ] , \sin [ \tau/R ] \right) .
\eeq
Here, the arbitrary length scale $R$ was  introduced in order to keep the argument of the trigonometric functions dimensionless. In this way, $\tau$ becomes a time coordinate with periodicity $\tau \sim \tau + 2 \pi R$\,. In order to work in the replicated geometry, relevant to the evaluation of the \cres, we extend the range of the time coordinate $\tau$ such that it is periodic with period $\tau \sim \tau + 2 \pi R  n$. This of course results in a conical excess at $\rho=0$. Later on, we  perform a Weyl transformation which turns our flat space to $S^1\times H^{d-1},$ where $R$ is the curvature radius of the hyperboloid. 
Since $C_D$ is a general property of the CFT, it does not depend on $R$.\footnote{In cases where the planar defect originates from a spherical one via a coordinate transformation, see  \eg section 2.2 of \cite{Belin:2013uta}, $R$ is the radius of the original sphere.} 
The periodicity of the Euclidean time coordinate introduces an effective temperature into our system $T\equiv T_0/n = (2\pi n\, R)^{-1}$, as we will discuss in detail in section \ref{sect-holo_setup}. For now, let us just point out that this fictitious temperature is merely a tool in our calculation, and we are still computing the entanglement entropy of a planer (or spherical) subregion of the CFT vacuum state.

For a defect deformed in an orthogonal direction by a function $f^a(y^i)$, cf. eq.~\eqref{deformation}, we can construct adapted coordinates (similar to Gaussian normal coordinates) in an expansion in the distance $\rho$ from the entangling surface.
The new coordinates are related to the old ones via the transformation
\beq
\begin{cases}
x'^a =  x^a  - f^a (y^i) - \frac{1}{d-2} \le x^a K^b x_b - \frac{1}{2} K^a x^2 \ri + \mathcal{O} (\rho^4) \\
y'^i = y^i +  \del_i f^a (y^i) \, x_a  - \frac{1}{2(d-2)} x^2 \del^i K^a x_a + \mathcal{O} (\rho^5) \, ,
\end{cases}
\label{eq:change_coordinates_adapted}
\eeq
where we defined the trace and the traceless parts of the extrinsic curvature as follows
\beq
K^a \equiv (K^a)^i_{\; i} \, , \qquad
\tilde{K}^a_{ij} \equiv K^a_{ij} - \frac{K^a}{d-2} \delta_{ij} \, ,
\label{eq:definition_extrinsic_curvature}
\eeq
and the extrinsic curvature is related to the profile of the deformation as
\begin{equation}\label{extcurvf}
    K^a_{ij} = - \del_i \del_j f^a + \mathcal{O}(f^2)\,.
\end{equation}
In terms of the new cylindrical coordinates $x'^a = \rho' \left( \cos [ \tau'/R ] , \sin [ \tau'/R ] \right)$,
the metric becomes\footnote{Here and in the following, we neglect the primes in order to simplify the notation, but the metric and the tensors are understood to be evaluated in the new adapted coordinate system.}
\beq
\begin{aligned}
ds^2 & =
 \left(1+ \tfrac{2 K^c x_c}{d-2}  \right)\left( \frac{\rho^2}{R^2} \, d\tau^2
+  d\rho^2 + [ \delta_{ij} +2  \tilde{K}^a_{ij}x_a ] dy^i dy^j + \tfrac{4}{d-2}   \partial_i  K^b\, x_b \rho  d\rho dy^i\right) + \dots\,.
\label{bdy_metric}
\end{aligned}
\eeq
In the above expression the dots stand for higher order terms in the distance $\rho$ from the entangling surface (see eq.~(2.17) of \cite{Bianchi:2016xvf} for a full specification of the orders in $\rho$ which were neglected in this expression). Further, note that we work at first order in the deformation $f^a$. It is remarkable that in the above metric the explicit dependence on the function $f^a$ dropped, and we only remain with dependence on $f^a$ via the extrinsic curvature \eqref{extcurvf}. The reason for this cancellation is explained below eq.~\eqref{eq:conformal_rescaling_adapted_coordinates}.

While in dimensions $d \geq 4$, it is sufficient to consider the traceless part of the extrinsic curvature and set $K^a=0$ to extract $C_D,$ this cannot be done in $d=3$, where the traceless part of the extrinsic curvature vanishes identically. 
Therefore, we keep $K^a\neq0$ in our expressions
which allows us to treat the cases $d \geq 3$ all at the same time. Of course, in $d=2$ the entangling surface consists of discrete points and so it has no extrinsic curvature and its shape cannot be deformed.

Next, we apply a Weyl transformations with conformal factor
\beq
\Omega = \Omega_1 \Omega_2  \equiv \le 1- \frac{K^a x_a}{d-2} \ri  \frac{R}{\rho}  \, .
\label{eq:conformal_rescaling_adapted_coordinates}
\eeq
The first rescaling $\Omega_1$ simply eliminates the overall factor in eq.~\eqref{bdy_metric}, and it is responsible for making the deformation $f^a$ enter the metric only via the combinations $\tilde{K}^a_{ij}$ and $\del_i K^a .$ 
This fact can be explained as follows.
In the special case when $f^a$ implements a conformal transformation, the change of coordinates to the adapted coordinate system introduced in eq.~\eqref{eq:change_coordinates_adapted} is the inverse transformation. 
Since the extrinsic curvature for a spherical defect (which is conformally related to the planar case) satisfies $\tilde{K}^a_{ij} = \del_i K^a =0,$ then we correctly obtain that the metric after the conformal rescaling $\Omega_1$ becomes flat when $f^a$ maps the planar defect to a sphere.
The second Weyl rescaling $\Omega_2 = R / \rho $ is introduced for convenience, in view of the holographic computation in section \ref{sect-holo_setup}.
As earlier, the factor of $R$ in the numerator was introduced to preserve the length dimensions of the infinitesimal line element. 
After the conformal transformation in eq.~\eqref{eq:conformal_rescaling_adapted_coordinates} is implemented, the metric becomes\footnote{Here too, the dots represent higher orders in $\rho$. The explicit expression for the orders which have been neglected can be found in eq.~(2.19) of \cite{Bianchi:2016xvf}.}
\begin{equation}\label{eq:metricdefhyp}
\begin{split}
ds^2 & =
   d\tau^2
+ \frac{R^2}{\rho^2} \left( d\rho^2 + [ \delta_{ij} +2 \, \tilde{K}^a_{ij}x_a ] dy^i dy^j + \frac{4}{d-2}   \partial_i \, K^b x_b \, \rho  d\rho dy^i \right) + \dots \, .
\end{split}
\end{equation}
This is a deformation of the hyperboloid $S^1 \times H^{d-1}$, at first order in the extrinsic curvature and for small values of the radial distance $\rho$ to the defect.
The curvature radius of the deformed hyperbolic space $H^{d-1}$ is $R$, see discussion below eq.~\eqref{cylindcoord1}.

Under the conformal rescaling \eqref{eq:conformal_rescaling_adapted_coordinates},
the one-point functions of the stress tensor and current transform as\footnote{The conformal factor is given by $\Omega^{\frac{s-\Delta}{2}},$ where $s$ is the spin of the field and $\Delta$ its scaling dimension.
For the stress tensor we have $(s=2, \Delta=d),$ while for the gauge current we use $(s=1, \Delta= d-1).$}
\beq
\langle \tilde{T}_{\mu\nu} \rangle_{n,\mu,f \Sigma} = \Omega^{2-d} \langle T_{\mu\nu} \rangle_{n,\mu,f \Sigma} + \mathcal{A}_{\mu\nu} \, ,
\label{eq:anomalous_transf_T}
\eeq
\beq
\langle \tilde{J}_{\mu} \rangle_{n,\mu,f \Sigma} = \Omega^{2-d} \langle J_{\mu} \rangle_{n,\mu,f \Sigma}  \, ,
\label{eq:anomalous_transf_J}
\eeq
where $\mathcal{A}_{\mu\nu}$ is an anomalous contribution, generalizing the Schwarzian derivative to the higher-dimensional case. 
In these expressions, the correlation functions on the right-hand side are taken in the presence of the deformed defect in the adapted coordinate system, while those on the left-hand side are evaluated in the deformed hyperboloid background \eqref{eq:metricdefhyp} after the conformal rescaling. 

The anomalous contribution $\mathcal{A}_{\mu\nu}$ in eq.~\eqref{eq:anomalous_transf_T}
does not depend on the number of replicas $n$ and on the chemical potential $\mu$.
The reason for the former is that locally the $n-$fold cover is identical to the original manifold, and therefore the anomaly functional, being a local quantity, is independent of the Rényi index.
Furthermore, the gauge invariance of the Weyl anomaly guarantees that there is no explicit dependence on the chemical potential.\footnote{The change in the one-point function of the stress tensor due to the Weyl anomaly was computed in \cite{PhysRevD.16.1712,Herzog:2013ed} for conformally flat spacetimes. In the presence of a gauge connection, the Weyl anomaly contains an additional term proportional to $F^2,$ which is gauge-invariant \cite{Duff:1993wm}.}
Therefore, we can extract the anomalous contribution by evaluating eq.~\eqref{eq:anomalous_transf_T} for
$n=1$ and $\mu=0$, which corresponds to a case without any 
defect where the expectation value of the stress tensor vanishes, \ie $\langle T_{\mu\nu} \rangle_{(n=1, \mu=0)} =0$
and therefore 
$\mathcal{A}_{\mu\nu} = \langle \tilde{T}_{\mu\nu} \rangle_{(n=1, \mu=0)}$.
Evaluating the term $\Omega^{2-d} \langle T_{\mu\nu} \rangle_{n,\mu,f \Sigma}$ on the right hand side of eq.~\eqref{eq:anomalous_transf_T}, we find
\begin{equation}\label{Tgoodframe}
\begin{aligned}
\langle \widetilde T_{ab} (x) \rangle_{n,\mu,f \Sigma}
& = R^{2-d} \, \frac{g_n (\mu)}{\rho^2}  \left((d-1) \delta_{ab} - d \frac{ x_a x_b}{\rho^2}\right) +\ldots
\, ,
\\
\langle \widetilde T_{ai} (x) \rangle_{n,\mu,f \Sigma}
& = R^{2-d} \, \frac{x_a x_b }{\rho^2}\del_i K^b \frac{k_n (\mu)}{d-2}  + \ldots \,  ,
\\
\langle \widetilde T_{ij} (x) \rangle_{n,\mu,f \Sigma}
& = R^{2-d} \, \frac{1}{\rho^2} \left( -g_n (\mu) \delta_{ij} + k_n (\mu) \tilde K^a_{ij} x_a \right) +\ldots \, ,
\end{aligned}
\end{equation}
where 
\begin{align}
\label{eq:CFToutputkn}
k_n (\mu) -k_1 (0) &=\frac{(d-1)   \Gamma \left(\frac{d}{2}-1\right)\pi ^{\frac{d}{2}-2}}{2\Gamma (d+1)}\frac{C_D}{n}-\frac{3d-4}{d-2} \frac{h_n (\mu)}{2\pi n} \, , \quad  &
g_n (\mu) -g_1 (0)&=\frac{h_n (\mu)}{2\pi n}\,.
\end{align}
As before, the ellipses denote the higher-order terms in $\rho$ which we neglected.

Following eq.~\eqref{eq:anomalous_transf_J}, we also find the transformed  current, whose one-point function 
(up to  first order in the deformation $f^a$ and at leading order in $\rho$) reads
\beq
\langle \tilde{J}_i \rangle_{n,\mu,f \Sigma} = 0 \, , \qquad
\langle \tilde{J}_a \rangle_{n,\mu,f \Sigma} = R^{2-d} \, \frac{i a_n (\mu)}{ 2 \pi n} \, \frac{\epsilon_{ab} x^b}{\rho^2}\, ,
\label{eq:deformed_dual_current}
\eeq
where in order to obtain this result we had to use  the two-dimensional tensor identity $\epsilon_{bc} x^c K^a x_a - x^a x_a \epsilon_{bc} K^c 
- \epsilon_{ac} K^a x^c x_b = 0$ which resulted in the non trivial cancellation of all the extrinsic curvature contributions to $J_a$. 
An important consequence of the cancellation of the linear order terms in the extrinsic curvature  
is that on the holographic side, we do not need to introduce a deformation of the dual gauge field, but we only deform the metric solution.

\subsection{Supersymmetric R\'enyi entropy}
\label{sec:SUSY_Renyi_entropy_QFT}

Explicit examples of conformal field theories with a global symmetry group are very common in supersymmetric theories. The superconformal group in any dimension $2\leq d \leq 6$ always includes an R-symmetry group under which the supercharges are charged. Larger amount of supersymmetry corresponds to higher-dimensional R-symmetry groups. One can select a $U(1)$ subgroup of the global R-symmetry and switch on the associated background gauge field, giving rise to the Dirac sheet we introduced above in section \ref{sect-replica_twist}. 
In this case, the generalized twist operator $\tilde\tau_n$ is a superconformal defect, i.e. it preserves part of the original supersymmetry. For sufficiently large supersymmetry, one can use supersymmetric localization to compute its spherical (or planar) expectation value and consequently the supersymmetric R\'enyi entropy. Exact results for the spherical supersymmetric R\'enyi entropy are available for $\mathcal{N}\geq2$ SCFTs in three dimensions, $\mathcal{N}\geq2$ SCFTs in four dimensions and $\mathcal{N}=1$ SCFTs in five dimensions \cite{Nishioka:2013haa,Hama:2014iea,Nishioka:2016guu}. From a bottom up holographic perspective, supersymmetry is implemented by requiring a specific relation between the charge and the mass of the hyperbolic black hole and this will be addressed in section \ref{sect-susy_solution}. Here we focus on the field-theoretical expectations.

An important feature of SCFTs in any dimensions is that the R-symmetry current and the stress tensor operator belong to the same supermultiplet. Using supersymmetric Ward identities under the preserved supersymmetry one can show that the magnetic response $a_n(\mu)$ is proportional to the conformal weight $h_n(\mu)$. The precise relation depends on the spacetime dimension and on which U(1) subgroup of the full R-symmetry group is gauged. An explicit example was given in \cite{Bianchi:2019sxz} for $\mathcal{N}\geq1$ SCFTs  in four dimensions. For $\mathcal{N}=1$, for instance, the R-symmetry is just U(1) and the stress tensor multiplet contains the U(1) R-symmetry current $J^{\mu}$, the supersymmetry currents $\mathcal{J}^{\mu \alpha}$, $\tilde{\mathcal{J}}^{\mu}_{\dot \alpha}$ and the stress tensor $T^{\mu\nu}$. Imposing that the supersymmetry variation of the fermionic one-point functions $\langle\delta_{\text{SUSY}}\mathcal{J}^{\mu \alpha}\rangle$ and $ \langle\delta_{\text{SUSY}}\tilde{\mathcal{J}}^{\mu}_{\dot \alpha}\rangle$ vanishes, one finds 
\beq\label{anhnrelsusy}
a_n(\mu) \propto h_n(\mu) \, .
\eeq
The precise relation between the two depends on the precise form of the current that is coupled to the background gauge field (for extended supersymmetry one has to choose a $U(1)$ subgroup of the total R-symmetry).  
Remarkably, in section \ref{sect-susy_solution} we will indeed find a proportionality relation between $a_n(\mu)$ and $h_n(\mu)$ which holds for supersymmetric R\'enyi entropies in any dimension.

A more involved implication of superconformal symmetry is the relation between the shape deformation of the entangling surface and the geometric deformation of the background, \ie the theory-independent proportionality relation \eqref{eq:conj_CD} between $C_D(\mu,n)$ and $h_n(\mu)$.  
A first evidence for the existence of such a relation appeared in \cite{Lewkowycz:2013laa,Fiol:2015spa} in the context of supersymmetric Wilson lines. For general, not necessarily supersymmetric theories, the authors of \cite{Bianchi:2015liz}  observed that the same relation holds 
at leading non-trivial order in an expansion around $n=1$. Away from $n=1$, \ie for general R\'enyi index, the relation \eqref{eq:conj_CD} holds for free field theories \cite{Bianchi:2021snj}, but it
 does not hold in holographic theories \cite{Dong:2016wcf,Bianchi:2016xvf}. This is in contrast to the case of superconformal theories, where it is believed that the relation is valid in general. In fact, the relation was proven for any superconformal defect in four dimensions \cite{Bianchi:2018zpb,Bianchi:2019sxz}. 
The derivation is based on the idea that using superconformal Ward identities, 
one can relate the correlator $\langle T_{\mu\nu}D_a \rangle$ to a correlator of operators with lower spin. While the correlator $\langle T_{\mu\nu}D_a \rangle$ is fixed in terms of two independent coefficients $C_D(n,\mu)$ and $h_n(\mu)$ \cite{Billo:2016cpy}, the lower-spin correlator has a reduced number of kinematic structures and is therefore fixed in terms of a single coefficient. This implies that the  relation \eqref{eq:conj_CD} between $C_D$ and $h_n$ should hold for superconformal defects in four dimensions and in particular for the supersymmetric R\'enyi entropies. We refer the reader to \cite{Bianchi:2018zpb,Bianchi:2019sxz} for the full details of the derivation. In the next section, we provide holographic evidence that the relation holds in any dimension, formally including cases with $d>6$ where there is no superconformal group and it is not even clear how a supersymmetric defect should be defined.

\section{The holographic story} 
\label{sec:holo}

In this section, we show how to compute $C_D$ in holographic theories.
We introduce the holographic setting in section \ref{sect-holo_setup}, which consists of a charged hyperbolic black hole in Einstein-Maxwell gravity. We also discuss how to tune the black hole solution to become supersymmetric.
In section \ref{sec:deformed_background}, we show how to deform the gravitational solution to account for shape deformations of the entangling surface. We then compute the expectation value of the stress tensor and the current in the deformed background using holographic renormalization techniques in section \ref{sect-holo_renormalization}. Finally, in sections \ref{sect-analytic} and \ref{sect-numerical}, we present analytical and numerical results for $C_D$.

\subsection{Holographic setup}
\label{sect-holo_setup}

Our previous discussion revolved around the charged R\'enyi entropy corresponding to a planar (or spherical) entangling surface in a constant time slice of the vacuum state of a $d$--dimensional CFT. As explained in \cite{Casini:2011kv,Hung:2011nu,Belin:2013uta}, a conformal transformation can be used to map the causal developments of the above regions 
to the hyperbolic cylinder $S^1 \times H^{d-1},$ where the Euclidean time coordinate $\tau$ becomes periodic on the Euclidean circle with period $\tau \rightarrow\tau + 2 \pi R$.\footnote{Recall that for the planar entangling surface $R$ is an arbitrary scale, which we introduced in order to give dimensions of length to the time coordinates and dimensions of inverse length squared to the curvature of the hyperboloid, as in \cite{Belin:2013uta}. For a spherical entangling surface, $R$ is the radius of the sphere. Of course, $C_D$ will not depend on this choice. See discussion below eq.~\eqref{cylindcoord1}. }
Under this map, the original reduced density matrix  describing a subregion of the CFT vacuum
is mapped to a thermal density matrix in the hyperbolic background with temperature $T_0= (2 \pi R)^{-1}$. 
Since we interpreted the Rényi entropies in terms of a partition function in an $n$-fold cover of the original spacetime, see  section \ref{sect-replica_twist}, applying  the same conformal transformation  to the replicated space results in a Euclidean time coordinate with periodicity 
$\tau \sim \tau + 2 \pi R\, n,$ and hence a thermal state with temperature of $T_0/n= (2 \pi R \,n)^{-1}$. The R\'enyi entropy can then be related to the thermal partition function with this temperature.

In a holographic CFT, 
we can 
evaluate the thermal partition function on the hyperbolic cylinder in terms of the horizon entropy of a hyperbolic AdS black hole solution see, \eg \cite{Casini:2011kv,Hung:2011nu,Belin:2013uta,Galante:2013wta}.
In the presence of a background gauge field, like the one introduced in our discussion of charged R\'enyi entropies below eq.~\eqref{eq:general_replica_trick}, one must further supplement the holographic setup with a background gauge field. Then, the gravitational setup
consists of a charged hyperbolic black hole solution to Einstein-Maxwell gravity \cite{Belin:2013uta}.

\subsubsection{Hyperbolic black holes in  Einstein-Maxwell gravity}
\label{sect-Einstein_Maxwell_gravity}

We consider the (Euclidean)
Einstein-Maxwell action in $d+1$ dimensions \cite{Chamblin:1999tk,Belin:2013uta}, 
\beq
I_{\rm EM} = - \frac{1}{2 \ell_P^{d-1}} 
\int d^{d+1} x \, \sqrt{g} \le \mathcal{R} + \frac{d(d-1)}{L^2} - \frac{\ell_*^2}{4} F_{\mu\nu} F^{\mu\nu}  \ri \, ,
\label{eq:EM_action}
\eeq
where $\ell_{\rm P}$ is the Planck length, $L$ is the curvature scale of AdS and $\ell_*$ is a coupling constant of the gauge field.  
This action admits charged black hole solutions, whose metric is given by
\beq
ds^2 =  G(r) \frac{L^2}{R^2} d \tau^2 + \frac{dr^2}{G(r)} + r^2 d\Sigma^2_{d-1} \, ,
\label{eq:metric_EM}
\eeq
where $\tau$ is the Euclidean time coordinate and $d \Sigma^2_{d+1} = du^2 + \sinh^2 u \, d\Omega^2_{d-1} $ is the metric on the hyperbolic space $H^{d-1}$ with unit curvature radius. In the above expression, we used the blackening factor which reads 
\beq
G(r) = \frac{r^2}{L^2} -1 - \frac{M}{r^{d-2}} + \frac{Q^2}{r^{2(d-2)}} \, ,
\label{eq:blackening_factor_Einstein_Maxwell}
\eeq
where $M$ and $Q$ are the mass and charge of the black hole, respectively. 
The metric has a horizon $r_h$, defined as the largest root of $G(r_h)=0$. This can be used to express the mass $M$ in terms of the charge and the horizon radius of the black hole,
\beq
M= \frac{r_h^{d-2}}{L^2} \le r_h^2 - L^2 \ri + \frac{Q^2}{r_h^{d-2}}
\label{eq:relation_mass_charge} \, .
\eeq
The gauge field solving the Einstein-Maxwell equations is\footnote{Note that in Euclidean signature the gauge field becomes imaginary.} 
\beq
A = -i \le  \sqrt{\frac{2(d-1)}{d-2}}  \frac{L Q}{R \ell_* \, r^{d-2}} -  \frac{\mu}{2 \pi R} \ri d\tau \, ,
\label{eq:zeroth_order_solution_gauge}
\eeq  
where the chemical potential $\mu$ is fixed by requiring that the gauge field vanishes at the horizon, \ie
\beq
\mu = 2 \pi \sqrt{\frac{2(d-1)}{d-2}}  \frac{L Q}{\ell_* \, r_h^{d-2}}  \, .
\label{eq:chemical_potential_and_charge}
\eeq

The Hawking temperature of the black hole can be expressed as
\beq
T(r_h, \mu) = \frac{T_0}{2} L \, G'(r_h) = 
\frac{T_0  L}{2 r_h}  \left[ d \, \frac{r_h^2}{L^2} - (d-2) - \frac{(d-2)^2}{2(d-1)} \le \frac{\mu \ell_*}{2 \pi L} \ri^2  \right]  \, , \label{temperature}
\eeq
where $T_0 = (2\pi R)^{-1}$ is the temperature introduced via the conformal mapping
at the beginning of section \ref{sect-holo_setup}. In what follows, it will be useful to work in terms of a dimensionless horizon radius, $x \equiv r_h/L$. 
As explained above, the R\'enyi entropies are obtained by studying a black hole with temperature  $T_0/n$. This requirement fixes the horizon radius to be 
\beq
x_n = \frac{1}{d \, n} + \sqrt{\frac{1}{d^2 n^2} + \frac{d-2}{d} +  \frac{(d-2)^2}{2d(d-1)} \le \frac{\mu \ell_*}{2 \pi L} \ri^2 } \, .
\label{eq:definition_xn}
\eeq
We can consider both real or imaginary chemical potential.
The latter case simply corresponds to analytically continuing  $\mu \to i \mu_{\rm E}$ and $Q \to i Q_{\rm E},$ where $\mu_{\rm E}, Q_{\rm E} \in \mathbb{R}.$
The requirement that the horizon radius $x_n$ is real 
bounds the Euclidean chemical potential from above by
\beq
\mu_E^2 \leq \frac{8 \pi^2 (d-1)}{d-2} \le \frac{L}{\ell_*} \ri^2 
\le 1 + \frac{1}{d(d-2)n^2} \ri \, .
\label{eq:domain_mu_imaginary}
\eeq
In order to compute the Rényi entropy in a given charge sector of the theory,
 we need to determine the partition function, see eq.~\eqref{symmetryresolvedRN2}.
In the grand-canonical ensemble, the partition function is determined from the grand potential $\mathcal{G}$ according to  $\mathcal{G}= - T \log Z(\mu).$ By evaluating the renormalized on-shell action, see appendix \ref{app:on_shell_action}, we obtain that 
\beq
\mathcal{G} = - \frac{V_{\Sigma} r_h^{d-2}}{2 \ell_{\mathrm{P}}^{d-1}} 
\left[   1+ \frac{d-2}{2(d-1)} \le \frac{\mu \ell_*}{2 \pi L} \ri^2  
+ \frac{r_h^2}{L^2}
\right] \, ,
\label{eq:Gibbs_free_energy}
\eeq
where $V_{\Sigma} \equiv \int_{H^{d-1}} d \Sigma_{d-1} $ denotes the dimensionless volume of the hyperbolic space $H^{d-1},$ regulated with the introduction of a UV cutoff, see \cite{Hung:2011nu}.\footnote{The divergence in the volume of the hyperbolic hyperplane is a reincarnation of the standard short-distance divergences in R\'enyi entropies.}   
Using eq.~\eqref{intro:partfunmu}, we obtain the corresponding Rényi entropy in terms of the grand potential, 
\begin{equation}
S_n(\mu)=\frac{n}{1-n} \frac{1}{T_0} \left[ \mathcal{G}(T_0) -\mathcal{G} \le \frac{T_0}{n} \ri \right] \, ,
\label{eq:Renyi_from_grand_potential}
\end{equation}
and, using eq. \eqref{eq:Gibbs_free_energy}, this becomes
\begin{equation}
    S_n(\mu) = \pi V_\Sigma \le \frac{L}{\ell_{\mathrm{P}}} \ri^{d-1}
    \frac{n}{n-1}
\left[  \le  1+ \frac{d-2}{2(d-1)} \le \frac{\mu \ell_*}{2 \pi L} \ri^2 \ri  \le x_1^{d-2} -x_n^{d-2}  \ri + x_1^d - x_n^d 
\right] \, .
\end{equation}
Finally, in order to match our gravity calculations with the CFT ones, it is useful to present the expressions for the central charges $C_T$ and $C_V$ -- defined in  eq.~\eqref{eq:definition_CTCV} -- in terms of  gravitational quantities \cite{Hung:2011nu,Belin:2013uta},
\beq
C_T = \le \frac{L}{\ell_{\rm P}} \ri^{d-1}  \frac{\Gamma(d+2)}{\pi^{d/2}(d-1)\Gamma(d/2)}\, ,\qquad \quad
C_V =  \frac{\Gamma(d)}{2 \pi^{d/2} \Gamma \le \frac{d}{2} -1 \ri} 
\frac{\ell_*^2 L^{d-3}}{\ell_{\rm P}^{d-1}} \, .
\label{eq:CTCV_holography}
\eeq

\subsubsection{Supersymmetric solution}
\label{sect-susy_solution}
In 3+1 bulk dimensions the black hole solution with blackening factor \eqref{eq:blackening_factor_Einstein_Maxwell} can be embedded in $\mathcal{N}=2$ gauged supergravity \cite{Freedman:1976aw}, whose Killing spinor equation reads  \cite{Nishioka:2014mwa} 
\beq
\left[ \nabla_{\mu} + \frac{1}{2L} \Gamma_{\mu} - \frac{i}{L} A_{\mu} 
+ \frac{i}{4} F_{\nu\rho} \Gamma^{\nu\rho} \Gamma_{\mu}
\right] \zeta = 0 \, , \label{killing_eq}
\eeq
where $\Gamma_{a}=e^{\mu}_a \Gamma_{\mu}$ are the Dirac matrices in the local Lorentz frame satisfying the Clifford algebra $\lbrace \Gamma_a, \Gamma_b \rbrace = 2 \delta_{ab},$ their antisymmetric combination is $\Gamma_{ab} = \frac{1}{2} [\Gamma_{a}, \Gamma_{b} ]$ and $\zeta$ is the Weyl spinor parametrizing the SUSY transformation.
Eq.~\eqref{killing_eq} corresponds to imposing that the SUSY variation of the gravitino vanishes.
One can show that its solution preserves 1/2 of the total supersymmetry when  \cite{Alonso-Alberca:2000zeh} 
\beq
M = 2 i  Q \, . \label{susy_rel}
\eeq
While eq.~\eqref{susy_rel} is explicitly derived in 3+1 dimensions, it holds in arbitrary dimensions (see \eg \cite{Hosseini:2019and} for the (5+1)-dimensional case). In our analysis, we will refer to the supersymmetric solution whenever we impose the relation \eqref{susy_rel}, independently of the spacetime dimensions.
In the supersymmetric case, the global charge is associated to the R-symmetry of the underlying theory.

Using the identities \eqref{eq:relation_mass_charge} and \eqref{eq:chemical_potential_and_charge} together with the relation \eqref{susy_rel}, we can express the R-charge and the chemical potential entirely in terms of the horizon radius as 
\beq
Q = i \, r_h^{d-2} \le 1 - \frac{r_h}{L} \ri \, , \qquad
\mu = 2 \pi i \, \sqrt{\frac{2(d-1)}{d-2}} \frac{L}{\ell_*} \le 1 - \frac{r_h}{L} \ri  \, .
\eeq
Consequently, the temperature in  \eqref{temperature} becomes 
\beq
T = \frac{1}{2 \pi R} \left[ (d-1) \frac{r_h}{L} - (d-2)  \right] \, .
\eeq
Upon imposing the restriction $T=(2\pi R\, n)^{-1}$ we obtain
\beq
x_n =   \frac{(d-2)n+1}{(d-1)n} \, .
\label{eq:xn_SUSY}
\eeq
Given that $x_n$ is only a function of $n$, the mass, the charge and the chemical potential all become functions of only $n$, for example, we find that
\beq
\mu_{\text{SUSY}} = 2 \pi i \sqrt{\frac{2}{(d-1)(d-2)}} \frac{L}{\ell_*} \frac{n-1}{n} \, .
\label{eq:chemical_potential_SUSY}
\eeq
In the following section, we will work with a general chemical potential $\mu$. The SUSY case will be treated separately by imposing the relation   \eqref{eq:chemical_potential_SUSY}.

\subsection{Deformed background}
\label{sec:deformed_background}

So far we only discussed a flat (or spherical) entangling surface. Now, consider deforming the entangling surface. For an arbitrary shape deformation, the dual bulk geometry is not known. However, in the case without charge, for small deformations, a solution can be constructed in an expansion in the distance $\rho$ from the entangling surface \cite{Dong:2016wcf,Bianchi:2016xvf,Mezei:2014zla}. Here we extend this approach to include charge.

We consider an ansatz for the bulk metric such that when approaching the boundary, we obtain -- up to a conformal rescaling -- the boundary metric \eqref{eq:metricdefhyp}. Furthermore, since higher orders in the near  boundary expansion of the metric  encode the boundary stress tensor, we introduce unknown functions in front of the coefficients corresponding to the unknown function $k_n(\mu)$ in eq.~\eqref{Tgoodframe}. Our ansatz for the bulk metric then reads
\beq
\begin{aligned}
ds^2_{\rm bulk} & =  \frac{dr^2}{G(r)} 
+ G(r)  \frac{L^2}{R^2} d\tau^2 \\
& + \frac{r^2}{\rho^2} 
\le  d \rho^2 + \left[ \delta_{ij} + 2 k(r) \tilde{K}^a_{ij} x_a  \right] dy^i dy^j 
+ \frac{4}{d-2} v(r) \del_i K^b x_b \rho d\rho dy^i  \ri + \dots \, ,
\end{aligned}
\label{eq:bulk_metric_ansatz}
\eeq
where $G(r)$ is the blackening factor \eqref{eq:blackening_factor_Einstein_Maxwell} of the unperturbed black hole and the ellipses stand for higher-order terms in $\rho.$
This expansion is the same one as in the uncharged case \cite{Bianchi:2016xvf}, except that now, the unknown functions will depend on the chemical potential. 
The metric corresponds to a deformed version of the hyperbolic cylinder $S^1 \times H^{d-1}$ with curvature radius $R.$ 
The radial functions $k(r), v(r)$ encode the effect of the traceless and trace parts of the extrinsic curvature of the deformed  entangling surface on the bulk geometry. The requirement that 
close to the boundary we recover the boundary metric \eqref{eq:metricdefhyp} imposes 
 $k(r\rightarrow \infty) = v(r\rightarrow \infty) \to 1$.
 
In addition, as usual in holography, we require the solution to be smooth at the horizon $r=r_h .$
Notice that, in principle, 
the case $d=3$ should be treated separately, since the traceless part $\tilde{K}^a_{ij}$ of the extrinsic curvature vanishes and there is only one independent function $v(r)$. However, as we will explain soon, the two functions $k(r)$ and $v(r)$ are equal to each other and so eventually, the $d=3$ and $d>3$ cases can be treated together.

In order to study \cre,~we need to supplement the metric solution with a gauge field.
However, as shown in section \ref{sect-adapted_coord}, the one-point function of the dual conserved current is not modified at first order in the extrinsic curvature by the deformation of the flat (or spherical) defect, see eq.~\eqref{eq:deformed_dual_current}.
For this reason, it is enough to just consider the zeroth-order solution for the gauge field in eq.~\eqref{eq:zeroth_order_solution_gauge}. 
However, the final result for $C_D$ extracted via the boundary stress tensor will nevertheless be modified due to the presence of the charge, as we will see below.

Using the metric ansatz \eqref{eq:bulk_metric_ansatz} and the gauge field \eqref{eq:zeroth_order_solution_gauge}, we compute the Einstein equations at first order in the deformation of the flat entangling surface.
The equations can be simplified by using the Gauss-Codazzi equation $\del_k K^a_{ij}= \del_j K^a_{ik}$, which is also a direct consequence of eq.~\eqref{extcurvf}. 
For the function $v(r)$ we obtain 
the following second-order differential equation 
\beq
v''(r) + \frac{r G'(r) + (d-1) G(r)}{r G(r)} \, v'(r) - \frac{(d-3)L^2 G(r) + r^2}{L^2 r^2 G(r)^2} \, v(r) = 0 \, .
\label{eq:differential_equations_G}
\eeq
This equation is formally the same that was obtained in pure Einstein gravity \cite{Bianchi:2016xvf}, with the difference that now the blackening factor is given by  eq.~\eqref{eq:blackening_factor_Einstein_Maxwell}, which depends on the charge. Einstein's equations also yield the algebraic equation 
\beq
k(r) = v(r) \, .
\label{eq:equality_kv}
\eeq
\emph{En passant}, we observe that this identity provides a non-trivial consistency check of the metric ansatz \eqref{eq:bulk_metric_ansatz}, whose expansion should yield the boundary stress tensor, since the same coefficient $k_n$ appeared in the stress tensor \eqref{eq:CFToutputkn} in front of both the traceless and the trace parts of the extrinsic curvature.

We remind the reader that the case $d=3$ is special since there is only one independent function $v(r)$ entering the bulk metric.
However, one can easily check that this function also satisfies eq. \eqref{eq:differential_equations_G}.
This, together with the equality $v(r)=k(r)$ in higher dimensions means that we can consistently study solutions for all $d \geq 3$ together,  by solving eq. \eqref{eq:differential_equations_G}.

We still need to impose boundary conditions. It is straightforward to check that the general form of the Laurent expansion of $v(r)$ close to the boundary is not influenced by the presence of the gauge field.
In dimensions $d=3,4$ the expansions read 
\beq
\begin{aligned}
& d=3 \, , ~~~~\qquad  v(r)= k(r) = 1 - \frac{L^2}{2 r^2} + \frac{L^3}{r^3} \beta_n (\mu) + \mathcal{O}(r^{-4}) \, , & \\
& d=4 \, , ~~~~\qquad  v(r)= k(r) = 1 - \frac{L^2}{2 r^2} + \frac{L^4}{r^4} \beta_n (\mu) + \mathcal{O}(r^{-5}) \, , & 
\end{aligned}
\label{eq:asymptotic_expansion_k}
\eeq
while in a general number of dimensions, we have 
\beq
\begin{aligned}
& \text{general $d$} \, , \qquad  v(r)= k(r) = 1 - \frac{L^2}{2 r^2} + \cdots + \frac{L^d}{r^d} \beta_n (\mu) + \mathcal{O}(r^{-(d+1)}) \, , & 
\end{aligned}
\label{eq:asymptotic_expansion_k2}
\eeq
where $\beta_n (\mu)$ is the first free coefficient in the expansion, \ie it is not fixed by the boundary conditions. This coefficient will be essential to compute $C_D$, as we will show in section \ref{sect-holo_renormalization}. Note that we have written explicitly the $\beta_n$-dependence on the chemical potential $\mu$, since it plays an important role in the subtraction of the anomalous contribution discussed above eq.~\eqref{Tgoodframe}. 
We refer the reader to appendix \ref{app:diff_eq_higherd} for the details of the calculation in $d=5,6,7$.

Imposing regularity at the horizon, for any $d \geq 3$, we obtain a similar series expansion given by
\beq
k(r) = q_n (\mu) \le \frac{r}{L} - x_n \ri^{\frac{n}{2}} + \cdots \, ,
\label{eq:near_horizon_expansion}
\eeq
where $x_n$ is determined by eq.~\eqref{eq:definition_xn} and $q_n (\mu)$ is the other free coefficient that specifies the solution to the second-order differential equation.

\subsection{Holographic renormalization}
\label{sect-holo_renormalization}

The next step to find $C_D$ is to compute the holographic expectation value for the stress tensor in the deformed geometry. For this, we use the holographic renormalization method \cite{deHaro:2000vlm,Skenderis:2002wp}, which we quickly review here.
We start from the metric in Fefferman-Graham (FG) form \cite{Fefferman:2007rka} 
\beq
ds^2_{\rm bulk} = \frac{L^2}{z^2} \le dz^2 + \mathfrak{h}_{\mu\nu} (z,x) dx^{\mu} dx^{\nu} \ri \, ,
\label{eq:FG_gauge}
\eeq
where $z=L^2/r+\dots$ with the dots indicating higher orders in $1/r$ (the full expressions to go to the FG gauge are specified in appendix \ref{app:details_holo_ren}) and 
\beq
\mathfrak{h}_{\mu\nu} = \mathfrak{h}_{\mu\nu}^{(0)} (x) + z^2 \mathfrak{h}_{\mu\nu}^{(2)} (x) + \dots + z^d \le \mathfrak{h}_{\mu\nu}^{(d)} (x)
 +    \tilde{\mathfrak{h}}^{(d)}_{\mu\nu} \,\log z \ri
 + \dots  \, .
 \label{eq:metric_FG_expansion}
\eeq
The coefficient $\tilde{\mathfrak{h}}_{\mu\nu}^{(d)}$ only appears when $d$ is even. The expectation value of the boundary stress tensor is then given by 
\beq
\langle T_{\mu\nu} \rangle_{\tilde{H}_n} = 
\frac{d}{2} \le \frac{L}{\ell_{\mathrm{P}}} \ri^{d-1} \left(\frac{L}{R}\right)^{d-2} \mathfrak{h}^{(d)}_{\mu\nu} + \mathcal{X}_{\mu\nu} [ \mathfrak{h}^{(m)}_{\mu\nu}, \tilde{\mathfrak{h}}_{\mu\nu}^{(d)}]_{m < d} \, ,
\label{eq:holographic_stress_tensor}
\eeq
where the subscript $\tilde{H}_n$ indicates that we are computing the one-point function of the stress tensor in the deformed hyperboloid background -- see eq.~\eqref{eq:metricdefhyp}.
The term $\mathcal{X}_{\mu\nu}$ includes lower order $\mathfrak{h}^{(i)}$'s, which are local functions of the field theory source $\mathfrak{h}^{(0)}_{\mu\nu}$ and therefore are completely fixed by the boundary geometry.
The coefficient $\tilde{\mathfrak{h}}^{(d)}_{\mu\nu}$ is related to conformal anomalies and enters this term as well, but it is scheme dependent and therefore, can be eliminated from the one-point function by introducing an appropriate counterterm. All the contributions to $\mathcal{X}_{\mu\nu}$ will be independent of $n$ and $\mu$ and therefore, their effect will cancel in subtractions like \eqref{eq:CFToutputkn}. Therefore, to determine $h_n$ and $C_D$, we will only include the contribution of  $\mathfrak{h}_{\mu\nu}^{(d)}$ to the above equation and  ignore the rest.\footnote{The interested reader can find expressions for such terms in, \eg \cite{deHaro:2000vlm}.}

Applying the above prescription for the case without the deformation, \ie with extrinsic curvature set to zero in the bulk metric \eqref{eq:bulk_metric_ansatz} and the field theory stress-tensor \eqref{Tgoodframe}-\eqref{eq:CFToutputkn}, we are able to extract the conformal dimension of the twist operator 
\beq
h_n (\mu) = \pi n \le \frac{L}{\ell_{\mathrm{P}}} \ri^{d-1}
\left[  x_n^{d-2} (1- x_n^2) - \frac{d-2}{2(d-1)} \le \frac{\mu \ell_*}{2 \pi L} \ri^2 x_n^{d-2}  \right] \, .
\label{eq:weight_twist_operator_with_chemical_potential}
\eeq
which of course, matches with the one found in \cite{Belin:2013uta}.

\subsubsection{Holographic dictionary for the magnetic response}

The gauge field obeys a similar near-boundary expansion, which can be used to find the boundary value of the conserved current.
The general expansion reads \cite{Hartnoll:2009sz,Skenderis:2002wp,Bianchi:2001kw}
\beq
A_{\mu} (x, z) =  A_{\mu}^{(0)} + z^2 \, A_{\mu}^{(2)} + \dots 
+ z^{d-2} \le A_{\mu}^{(d-2)} +  \tilde{A}_{\mu}^{(d-2)} \, \log z  \ri + \dots  \, .
\eeq
The main difference with the metric expansion \eqref{eq:metric_FG_expansion} is that the undetermined term corresponds to the order $z^{d-2},$ as a consequence of the different scaling dimension of the dual current with respect to the energy-momentum tensor.
This asymptotic behaviour matches with the zeroth-order solution \eqref{eq:zeroth_order_solution_gauge} of the Einstein-Maxwell equations.
Working in the gauge $A_z = 0$, the 
 previous expansion determines a dual current 
\beq
\langle J_{\mu} \rangle_{\tilde{H}_n} = - \frac{d-2}{2} \left(\frac{L}{\ell_P}\right)^{d-1}
\left(\frac{\ell_* }{ L}\right)^2 \left(\frac{L}{R}\right)^{d-2} A_{\mu}^{(d-2)} \, .
\eeq
As mentioned at the beginning of the section, the gauge field solution \eqref{eq:zeroth_order_solution_gauge} receives no corrections at the first order in the deformation and leading order in $\rho$. This is consistent with the absence of extrinsic curvature corrections in the field theory  current in equation
\eqref{eq:deformed_dual_current}.
Using the above relation to the dual current fixes the magnetic response
\beq
a_n (\mu) = \frac{d-2}{2} \, \frac{\ell_*^2 L^{d-3}}{ \ell_{\rm P}^{d-1}} \,
n x_n^{d-2} \mu \, ,
\label{eq:an_from_holography}
\eeq
which is, of course, in agreement with the result for the undeformed case, obtained previously in \cite{Belin:2013uta}.

Note that in the supersymmetric case, where the identities \eqref{eq:xn_SUSY} 
and \eqref{eq:chemical_potential_SUSY} hold, the magnetic response becomes 
\beq\label{relationanhn}
a_n^{\text{SUSY}} (\mu) = i \frac{\ell_*}{L} \sqrt{\frac{(d-1)(d-2)}{2}} \, h_n (\mu) = i \sqrt{2d(d+1)} \sqrt{\frac{C_V}{C_T}} \, h_n (\mu) \, ,
\eeq
where $C_V$ and $C_T$ were defined in  eq.~\eqref{eq:CTCV_holography}. In a superconformal theory, the R-symmetry currents and the stress energy tensor belong to the same supermultiplet, therefore their vacuum two-point functions, $C_V$ and $C_T$, are related by superconformal Ward identities. The precise numerical factor depends on the specific $U(1)$ subgroup associated to the current $J_\mu$ in eq.~\eqref{eq:definition_CTCV} and our bottom-up holographic model is blind to this choice. In top-down examples, the Einstein Maxwell gravitational action can be derived from appropriate compactifications of 11d supergravity or of 10d type IIB supergravity. In that case, the ratio $\ell^*/L$ (and equivalently the ratio $C_V/C_T$) is fixed \cite{Chamblin:1999tk} (for instance $\ell^*/L=2$ for the compactification of $AdS_5\times S^5$ and $\ell^*/L=4$ for $AdS_4\times S^7$). As we mentioned in section \ref{sec:SUSY_Renyi_entropy_QFT}, a relation between $a_n^{\text{SUSY}} (\mu)$  and $h_n(\mu)$ is expected from supersymmetric Ward identities, but again the numerical factor depends on the choice of the current. What we find here, however, is that the ratio $a_n^{\text{SUSY}} (\mu)/h_n(\mu)$ is fully determined by $C_V/C_T$, which is a property of the CFT and, in particular, is certainly independent of $n$ and $\mu$. Notice also that the relation \eqref{relationanhn} is independent of the normalizations of the stress tensor and the current as it can be written in terms of the ratios $\frac{h_n(\mu)}{\sqrt{C_T}}$ and $\frac{a_n(\mu)}{\sqrt{C_V}}$.

\subsubsection{Holographic dictionary for $C_D$}

Using eqs.~\eqref{eq:FG_gauge} and \eqref{eq:holographic_stress_tensor}, we obtain a holographic prediction for the field theory coefficients defined in eq.~\eqref{Tgoodframe}, \ie
\beq
\begin{aligned}
& g_n (\mu) = - \le \frac{L}{\ell_{\mathrm{P}}} \ri^{d-1} \frac{1}{2} \left[ x_n^d + x_n^{d-2}
\le  \frac{d-2}{2(d-1)} \le \frac{\mu \ell_*}{2 \pi L} \ri^2 -1 \ri + g_0^{(d)}  \right] \, , & \\
& k_n  (\mu)= \le \frac{L}{\ell_{\mathrm{P}}} \ri^{d-1} \left[ x_n^d + x_n^{d-2} \le \frac{d-2}{2(d-1)} \le \frac{\mu \ell_*}{2 \pi L} \ri^2 -1 \ri + d \beta_n (\mu) + k_0^{(d)}  \right] \, . &
\end{aligned}
\label{eq:conformal_weight_twist_operator_from_FG_coordinates}
\eeq
Recall that $\beta_n(\mu)$ is the first undetermined coefficient in the Laurent expansion of the function $v(r)$ appearing in our metric, cf.~eq.~\eqref{eq:asymptotic_expansion_k2}. The above relations are valid in any dimension $d \geq 3$ and for any value of the chemical potential $\mu$. The coefficients $g_0^{(d)}$ and $k_0^{(d)}$ contain the anomalous contribution; they vanish in odd dimensions, and they do not depend on the R\'enyi index nor on the chemical potential in even dimensions.\footnote{The anomalous coefficients in even dimensions are $g_0^{(4)} = 1/4 , ~k_0^{(4)} = 3/4 $ in $d=4$ and $ g_0^{(6)} = - 3/8,~ k_0^{(6)} = 5/8 $ in $d=6.$
Their precise expression can be found by following the procedure outlined in appendix \ref{app:details_holo_ren} and including the anomalous contribution $\mathcal{X}_{\mu\nu}$ in eq.~\eqref{eq:holographic_stress_tensor}, which are specified  explicitly in eqs.~(3.15) and (3.16) of \cite{deHaro:2000vlm}.}
To determine $C_D$ and $h_n$ we only need differences of the coefficients $g_n,~ k_n,$ (cf.~eq.~\eqref{eq:CFToutputkn}) so these anomalous contributions will not play any role in the remaining of the computation.

We can now use eq.~\eqref{eq:CFToutputkn} to determine the conformal weight of the twist operator $h_n$ and the coefficient $C_D$. 
The former turns out to be the same expression determined for the undeformed case in eq.~\eqref{eq:weight_twist_operator_with_chemical_potential}; the latter reads 
\beq
C_D (n,\mu) = \frac{n \,  d \, \Gamma (d+1)}{(d-1) \pi^{d/2-2} \Gamma(d/2)}
\left[ (d-2) \le \frac{L}{\ell_{\mathrm{P}}} \ri^{d-1} \le \beta_n (\mu) - \beta_1 (0) \ri + \frac{h_n (\mu)}{2 \pi n} \right] \, .
\label{eq:holographic_equation_CD}
\eeq
Note that formally, this expression resembles the one obtained for the uncharged case. However, here the dependence on the charge enters through the constant $\beta_n(\mu)$ as well as through the $\mu$ dependence of $h_n(\mu)$. 
The remaining step is to compute $\beta_n(\mu)$. Once we have $\beta_n(\mu)$, it is straightforward to compute $C_D$ and check under which circumstances the conjecture in eq. \eqref{eq:conj_CD} holds.

\subsection{Analytic expansions}
\label{sect-analytic}

It is not possible to solve in full generality the differential equation \eqref{eq:differential_equations_G}, but we can solve it analytically, order-by-order in a double perturbative expansion around $n=1$ and $\mu=0.$ 
For this purpose, it is convenient to introduce the change of variables $\tilde{r} = (x_n L)/r, $ such that the range of the radial coordinate becomes compact.
The boundary is then located at $\tilde{r}=0,$ while the horizon sits at $\tilde{r}=1.$
The solution to the differential equation \eqref{eq:differential_equations_G} at order zero around $\mu=0$ and up to second order near $n=1$ was already found in \cite{Bianchi:2016xvf}.
Moving to higher orders in the chemical potential, one can similarly obtain solutions for the functions $v(r)$ and $k(r)$ in the double series expansion around $n=1$ and $\mu=0$.
These, in turn, determine $\beta_n (\mu)$ perturbatively,
\beq
\beta_n (\mu) = \sum_{A,B} \beta_{AB} (n-1)^A \mu^B  \, .
\label{eq:analytic_expansion_beta}
\eeq
Using 
eqs.~\eqref{eq:CTCV_holography}, \eqref{eq:weight_twist_operator_with_chemical_potential} and \eqref{eq:holographic_equation_CD}, we can then determine the analytic expansions for $C_D$ and the central charge $C_T.$
We present the analytic expansions in terms of the convenient combination
\beq
\tilde{\mu} \equiv \frac{\ell_*}{4 \pi L}  \,   \mu \, .
\label{eq:definition_mutilde}
\eeq
Below, we present the final results for $d=3,4$. We leave the details of the derivation and the results for the higher-dimensional cases to appendix \ref{app:diff_eq_higherd}.
We assume that the perturbative expansion is governed by a small parameter $\varepsilon$ such that $(n-1) = c_1 \varepsilon$ and $\tilde{\mu}^2 = c_2 \varepsilon $ for two arbitrary order-one constants $c_1, c_2.$
In the following, we present analytic expansions up to order $\varepsilon^2$.

In $d=3,$ we obtain for $C_D$ (conveniently normalized)
\beq
\frac{C_D}{C_T}\biggr|^{d=3} = 
- \frac{2}{3} \pi^2 \tilde{\mu}^2 
+ \frac{\pi^2}{2} (n-1) \le 1 - \frac{2 \log 2 +1}{6} \tilde{\mu}^2 \ri
- \frac{5 }{12} \pi^2  (n-1)^2 
+ \mathcal{O}(\varepsilon^3) \, .
\label{eq:analyitic_expansions_CD_over_CT_d3}
\eeq
The conjectured value of $C_D^{\rm conj}$ in eq.~\eqref{eq:conj_CD} reads
\beq
\frac{C_D^{\rm conj}}{C_T}\biggr|^{d=3} = - \frac{3}{4} \pi^2 \tilde{\mu}^2 
+ \frac{\pi^2}{2}  (n-1) \le 1 - \frac{3}{8} \tilde{\mu}^2 \ri
- \frac{7}{16} \pi^2  (n-1)^2 
+ \mathcal{O}(\varepsilon^3) \, .
\eeq
Subtracting the two expansions for $C_D$ and $C_D^{\rm{conj}},$ we obtain
\beq
\frac{C_D - C_D^{\rm conj}}{C_T}\biggr|^{d=3} = 
\frac{\pi^2}{12} \tilde{\mu}^2 
+ \frac{\pi^2}{2} \frac{5-8 \log 2}{24} \tilde{\mu}^2
+ \frac{\pi^2}{48}  (n-1)^2 
+ \mathcal{O}(\varepsilon^3) \, .
\label{eq:difference_CDCT_d3}
\eeq
Note, that the agreement between $C_D$ and $C_D^\text{conj}$ at first order in $(n-1)$ in the case without charge ($\tilde\mu=0$) was previously proven in broad generality in \cite{Faulkner:2015csl} and verified using holography in \cite{Bianchi:2016xvf}. 
We can also fix $n=1$, and expand around $\tilde{\mu}=0$. This will be useful to compare with our numerical results in the next section.
In $d=3$, we find 
\beq
\frac{C_D}{C_T}\Big|_{n=1}^{d=3} = 
- \frac{2}{3} \pi^2 \tilde{\mu}^2 + \frac{\pi^2}{540} \le 255 \log 2 - 248 \ri  \tilde{\mu}^4  
+ \mathcal{O} (\tilde\mu^6) \, ,
\label{eq:analyitic_expansions_CD_over_CT_d3_n1}
\eeq

We can perform analogous computations in $d=4$. For $C_D$, we find 
\beq
\frac{C_D}{C_T}\biggr|^{d=4} = 
- \frac{6}{5} \pi^2 \tilde{\mu}^2 
+ \frac{\pi^2}{5} \le 2 - \frac{\tilde{\mu}^2}{9} \ri (n-1) 
- \frac{11 \pi^2}{30}  (n-1)^2 
+ \mathcal{O}(\varepsilon^3)  \, ,
\label{eq:analyitic_expansions_CD_over_CT_d4}
\eeq
\beq
\frac{C_D^{\rm conj}}{C_T}\biggr|^{d=4} = - \frac{4}{3} \pi^2 \tilde{\mu}^2 
+ \frac{\pi^2}{5} \le 2 + \frac{4 \tilde{\mu}^2}{27} \ri (n-1) 
- \frac{17 \pi^2}{45}  (n-1)^2 
+ \mathcal{O}(\varepsilon^3)  \, ,
\eeq
and the difference
\beq
\frac{C_D - C_D^{\rm conj}}{C_T}\biggr|^{d=4} = 
\frac{2}{15} \pi^2 \tilde{\mu}^2 - \frac{7}{135} \pi^2 \tilde{\mu}^2 (n-1)
+ \frac{\pi^2}{90}  (n-1)^2 
+ \mathcal{O}(\varepsilon^3)  \, .
\label{eq:difference_CDCT_d4}
\eeq
In the uncharged case ($\tilde \mu=0$),
these expressions agree with the results of \cite{Bianchi:2016xvf} and in particular, 
the difference $C_D-C_D^{\rm conj}$ starts at order $(n-1)^2$, in agreement with the general proof of \cite{Faulkner:2015csl}.
The inclusion of the chemical potential is responsible for a discrepancy between $C_D$ and $C_D^{\rm conj}$ already when $n=1$, as can be seen from the first term.
This clearly shows that in the presence of a gauge field, the conjecture \eqref{eq:conj_CD} is violated even around $n=1.$
This pattern holds for any dimensions, as we report explicitly in appendix \ref{app:diff_eq_higherd}. 

Let us comment on the sign of $C_D$. For vanishing chemical potential, we recover the result of \cite{Bianchi:2016xvf}, \ie the displacement two-point function becomes negative for $n<1,$ violating unitarity (it is unclear whether a non-unitary defect interpretation remains valid in that regime). When the chemical potential is switched on, the zero of $C_D$ is no longer at $n=1$ and only for imaginary values $\tilde \mu^2<0$ it moves to the left, leaving a unitarity defect for $n\geq1$. 

We can also fix $n=1$, and expand around $\tilde{\mu}=0$. This will be useful to compare with our numerical results in the next section.
In $d=4,$ it turns out that we can reach a remarkable precision,
\beq
\frac{C_D}{C_T}\Big|_{n=1}^{d=4} = 
- \frac{6 \pi^2}{5} \tilde{\mu}^2
-\frac{503 \pi ^2}{540}  \tilde{\mu}^4
  +\frac{18011   \pi ^2 }{145800}  \tilde{\mu}^6
  -\frac{129908809 \pi ^2 }{1763596800}  \tilde{\mu}^8
  + \frac{739450117 \pi ^2 }{13604889600} \tilde{\mu}^{10} + \mathcal{O} (\tilde{\mu}^{12}) \, .
  \label{eq:analytic_result_d4_n1}
\eeq

\subsubsection{Supersymmetric case}

The supersymmetric case is special in that the chemical potential is completely fixed in terms of the number of replicas and the dimensionality of the spacetime, see eq.~\eqref{eq:chemical_potential_SUSY}.
For this reason, the series expansion looks different and the differential equation can be studied by expanding the solution around $n=1.$
We report here the results for $\beta_n$ in $d=3,4$,
\beq
\begin{aligned}
&  d=3 \, , \qquad \beta_n =  \frac{1}{6} (n-1) - \frac{1}{4} (n-1)^2 
+ \mathcal{O} (n-1)^3 \, , & \\
& d=4 \, , \qquad \beta_n = - \frac{1}{8} + \frac{1}{12} (n-1) - \frac{5}{36} (n-1)^2 
+ \mathcal{O} (n-1)^3 \, , & \\
\end{aligned}
\label{eq:susy_expansion_betan}
\eeq
which imply\footnote{In fact, these expansions can be obtained from the general expansions which appeared earlier by using the relation \eqref{eq:chemical_potential_SUSY}. However note that in this case the correct scaling is $\tilde \mu \sim (n-1) \sim \varepsilon$ and hence the mixed term $\tilde \mu^2 (n-1)$ will not contribute.}
\beq
\begin{aligned}
&  d=3 \, , \qquad \frac{C_D}{C_T} = \frac{C_D^{\rm conj}}{C_T} = \frac{\pi^2}{2} (n-1) - \frac{\pi^2}{4} (n-1)^2 + \mathcal{O}(n-1)^3 \, ,  & \\
& d=4 \, , \qquad \frac{C_D}{C_T} = \frac{C_D^{\rm conj}}{C_T} = \frac{2 \pi^2}{5} (n-1) - \frac{4 \pi^2}{15} (n-1)^2 + \mathcal{O}(n-1)^3 \, . & \\
\end{aligned}
\label{eq:susy_expansion_CDCT}
\eeq
As explicitly written, the relation $C_D = C_D^{\rm conj} $ holds at least to second order in perturbation theory. Moreover, in higher dimensions the perturbative expansion of the supersymmetric solution can be done at higher orders (see appendix \ref{app:diff_eq_higherd}), and the matching continues to hold order-by-order in the expansion. This contrasts with the uncharged (not-supersymmetric) case where the matching only holds to first order around $n=1$ \cite{Bianchi:2016xvf}. In section \ref{sect-numerical}, we will show that, indeed, in the supersymmetric case the conjecture holds (at least to numerical precision), for any value of $n$.

\subsection{Numerical results}
\label{sect-numerical}

In order to solve eq.~\eqref{eq:differential_equations_G} numerically, we employ a shooting method.
The two integration constants arise from the boundary conditions at the horizon and at infinity.
The former is parametrized by $q_n (\mu)$ in eq.~\eqref{eq:near_horizon_expansion}, while the latter by $\beta_n (\mu)$ in eq.~\eqref{eq:asymptotic_expansion_k2}.

Briefly, the numerical shooting method works as follows: we fix a specific value of the chemical potential, and then, for each value of $n$, the differential equation is solved numerically close to the horizon and close to the boundary.
The integration constants are determined by requiring that the two curves meet smoothly in the middle point.
After this numerical evaluation, we choose another fixed value of the chemical potential and we repeat the procedure.

\begin{figure}[h!]
\centering
\subfigure[]
{ \includegraphics[scale=0.82]{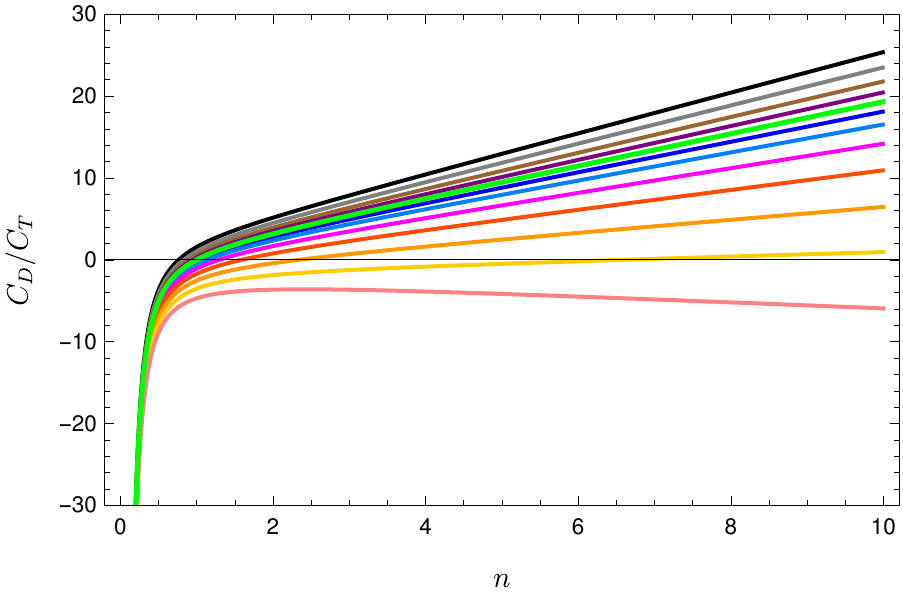}} 
\subfigure[]
{ \includegraphics[scale=0.82]{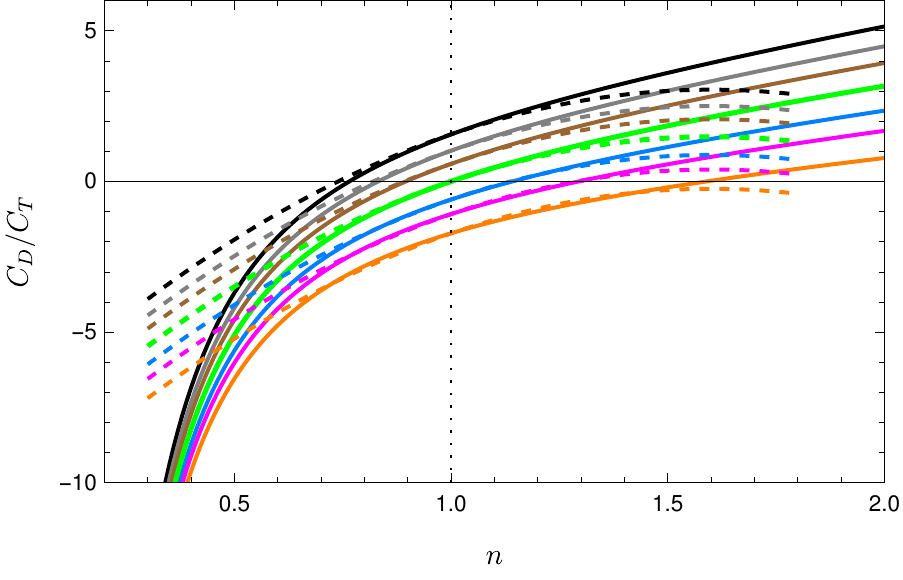}} \\
\caption{Plots of $C_D/C_T$, defined in eqs.~\eqref{eq:CTCV_holography} and \eqref{eq:holographic_equation_CD}, 
as a function of $n$ in $d=3.$ The colours correspond to different fixed values of the chemical potential $\tilde{\mu}$ in eq.~\eqref{eq:definition_mutilde} with $\ell_{*}/L=1.$ 
(a) The different curves correspond to $\tilde{\mu}^2 =\pm 0.04, \pm 0.09, \pm 0.16, \pm 0.25, 0.36, 0.49, 0.64$, with $\tilde{\mu}^2$ increasing from top to bottom. We highlight in green the curve corresponding to $\tilde{\mu}=0$. Note that $\tilde{\mu}^2$ can be negative, as the chemical potential can take imaginary values.
 (b) Zoom of the same plot around the region close to $n=1$. The dashed lines correspond to the analytic expansions in eq.~\eqref{eq:analyitic_expansions_CD_over_CT_d3}. Curves with the same colours correspond to the same chemical potential.}
\label{fig:CD3}
\end{figure}
\begin{figure}[h!]
\centering
\subfigure[]
{ \includegraphics[scale=0.82]{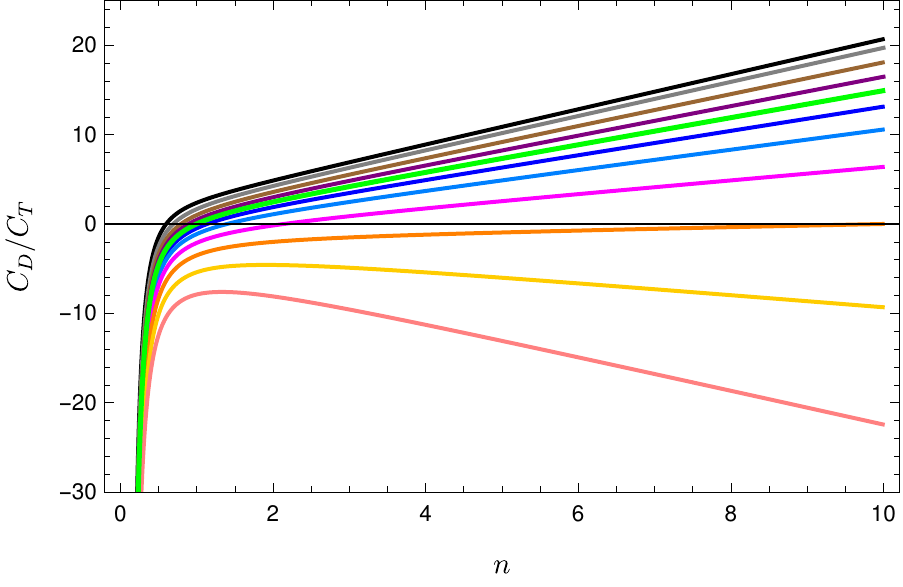}} 
\subfigure[]
{ \includegraphics[scale=0.82]{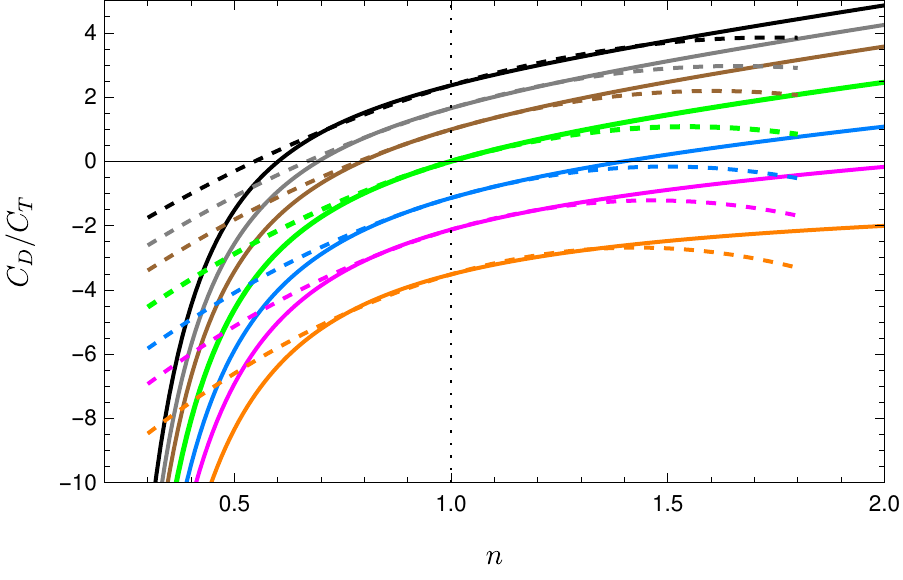}} \\
\caption{Plots of $C_D/C_T$, defined in eqs.~\eqref{eq:CTCV_holography} and \eqref{eq:holographic_equation_CD}, as a function of $n$ in $d=4.$ The colours correspond to different fixed values of the chemical potential $\tilde{\mu}$ in eq.~\eqref{eq:definition_mutilde} with $\ell_{*}/L=1.$ 
(a) The different curves correspond to $\tilde{\mu}^2 =\pm 0.04, \pm 0.09, \pm 0.16, \pm 0.25, 0.36, 0.49$, with $\tilde{\mu}^2$ increasing from top to bottom. We highlight in green the curve corresponding to $\tilde{\mu}=0$. Note that $\tilde{\mu}^2$ can be negative, as the chemical potential can take imaginary values.
 (b) Zoom of the same plot around the region close to $n=1$. The dashed lines correspond to the analytic expansions in eq.~\eqref{eq:analyitic_expansions_CD_over_CT_d4}. Curves with the same colours correspond to the same chemical potential.
}
\label{fig:CD4}
\end{figure}

We present results for $d=3$, $4$ in the main text, while leaving the higher dimensional cases for appendix \ref{app:diff_eq_higherd}. We first plot the value of $C_D$ as a function of $n$. Within each plot, the different coloured  curves correspond to different values of the chemical potential. This is shown in fig.~\ref{fig:CD3} for $d=3$ and in fig.~\ref{fig:CD4}, for $d=4$. Qualitatively, the results are similar in $d=3$, $4$. It is convenient to describe them in terms of $\tilde{\mu}^2$. Curves with $\tilde{\mu}^2 <0$, correspond to imaginary chemical potential, while curves with positive $\tilde{\mu}^2$ have real chemical potential. In all dimensions, the chemical potential squared increases when going from the top curve towards the bottom curve of each plot. 
An intermediate green curve, that corresponds to the uncharged case  ($\tilde{\mu}=0$), separates curves with imaginary and real chemical potential. This curve is the same one reported in \cite{Bianchi:2016xvf}. In all cases shown, $C_D$ follows a linear trend with respect to $n$ for large enough $n$. Close to $n=1$, for sufficiently small $\tilde{\mu}$, the numerical curves agree with the analytic expansions in the previous section. This is shown in the right panel of each figure, where the analytic curves are shown in dashed lines. Furthermore, the range of the Rényi index such that $C_D \geq 0$ -- which is expected for a unitary dCFT-- increases when considering imaginary chemical potentials. In particular, for imaginary chemical potential, $C_D$ is always positive for any R\'enyi index $n\geq 1$ (including, of course, all integer values of $n$); the case with imaginary chemical potential is the usual one which appears in the condensed matter literature, see for example the integration in  eq.~\eqref{symmetryresolvedRN2}; in this case, the chemical potential simply produces phases for charged fields as they go around the entangling surface.  However, as pointed out by \cite{Belin:2013uta}, in holography, one can also have a real value for the chemical potential.
We notice that, past a certain value of real $\tilde{\mu},$ the function $C_D$ becomes negative along all the range of the Rényi index.

\begin{figure}[h!]
\centering
\subfigure[]
{ \includegraphics[scale=0.82]{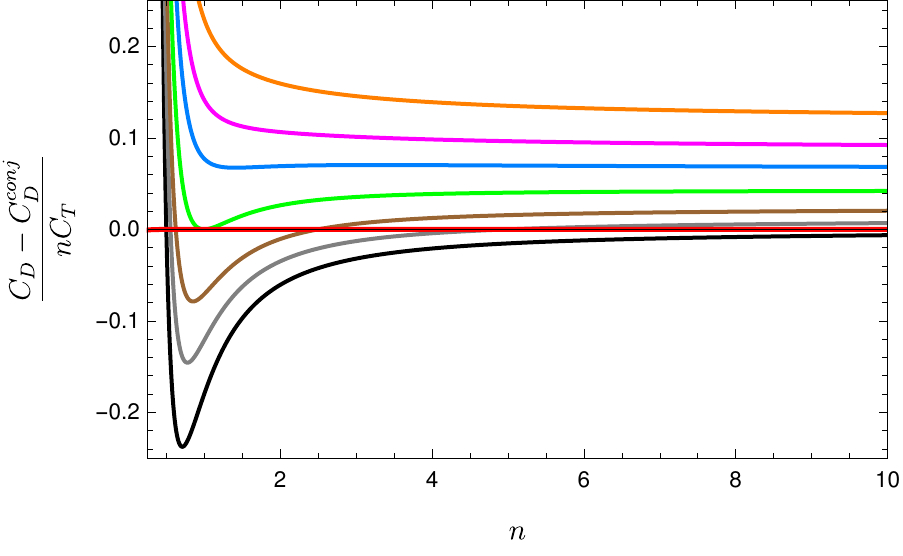}} 
\subfigure[]
{ \includegraphics[scale=0.82]{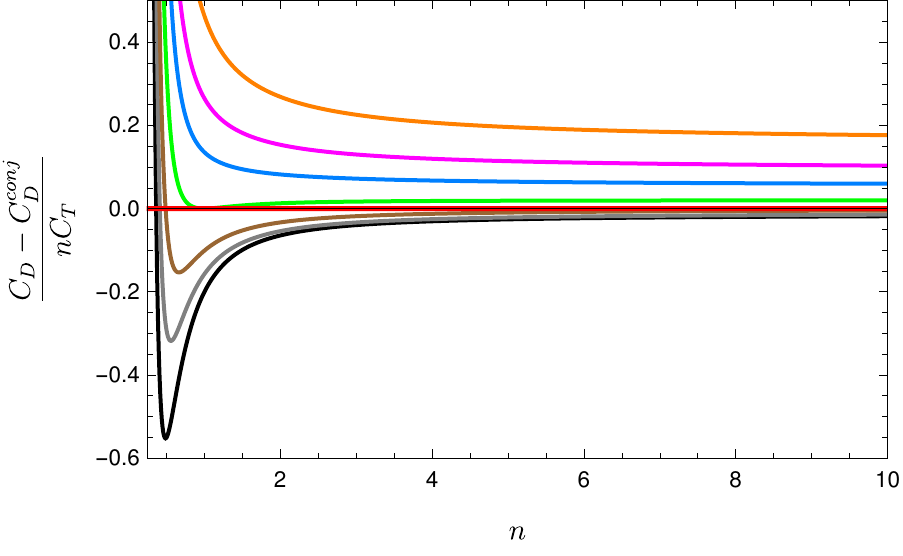}} \\
\caption{Relative difference $(C_D - C_D^{\rm conj})/(n C_T)$ as a function of $n$ for fixed chemical potential in $d=3$ (left) and $d=4$ (right). The different curves correspond to different chemical potentials, $\tilde{\mu}^2 = \pm 0.09, \pm 0.16, \pm 0.25$, with $\tilde{\mu}^2$ increasing from bottom to top (we set $\ell_{*}/L=1$). We plot the curve with $\tilde{\mu}=0$ in green. Note that $\tilde{\mu}^2$ can be negative, as the chemical potential can take imaginary values.
The red curve corresponds to evaluating the SUSY solution. As it can be seen, in this case $C_D$ is exactly $C_D^{\rm conj}$, at least up to numerical errors of the order $10^{-4}$ in $d = 3$ and $10^{-6}$ in $d = 4$.}
\label{fig:CD3_rel}
\end{figure}

In fig.~\ref{fig:CD3_rel}, we explicitly compare the numerical result with the conjectured value in eq. \eqref{eq:conj_CD}, both for $d=3$, $4$. It is convenient to plot the following difference,
\beq
\frac{C_D - C_D^{\rm conj}}{n C_T} \, ,
\eeq
where the factor of $n$ in the denominator is chosen in order to have a constant value for large Rényi index, and $C_T$ is the central charge, see eq.~\eqref{eq:definition_CTCV}.
Notice that curves with real chemical potential do not intersect the horizontal axis, while curves with imaginary chemical potential intersect it twice at two different values of $n$. The green curve, that corresponds to $\tilde{\mu}=0$, intersects at a single point given by $n=1$.  

In the plot, we observe that the conjectured result \eqref{eq:conj_CD} does not hold for generic values of the chemical potential. However, if we focus on the supersymmetric case, given by the red curve, then we see that the conjectured result holds (at least up to errors of the order $10^{-4}$ in $d=3$ and  $10^{-6}$ in $d=4$). It is interesting to note that the red curve is composed by the intersections of all the curves at imaginary chemical potential with the horizontal axis, plus the point at $n=1$ which comes from the intersection of the $\tilde{\mu}=0$ curve. The reason is that the chemical potential in the supersymmetric case is purely imaginary and a function of $n$, see eq.~\eqref{eq:chemical_potential_SUSY}.

Going back to the case of non-supersymmetric theories, we could try to quantify how badly is the conjecture \eqref{eq:conj_CD} violated at generic chemical potential.
One way to asses the level of violation is to look at the relative error $(C_D - C_D^{\rm conj})/C_D$.
However, it turns out that since the numerator and denominator have zeros at different values of $n,$ this function is divergent. Nevertheless, focusing on the regime $n>1$ and $\tilde\mu^2<0$ (which is the relevant regime for a unitary conformal field theory, see footnote \ref{physicalchem} and comments below equation \eqref{eq:difference_CDCT_d4}) we find that the conjecture is only mildly violated for a wide range of $n$. This is similar to what was found in the uncharged case in \cite{Bianchi:2016xvf}. For instance, for the values that we have studied numerically, the relative error approaches values smaller than $0.1$ for $n \geq 4$ and imaginary chemical potentials. On the other hand, for real chemical potential it seems that we can reach large violations.

\begin{figure}[h!]
\centering
\subfigure[]
{ \includegraphics[scale=0.82]{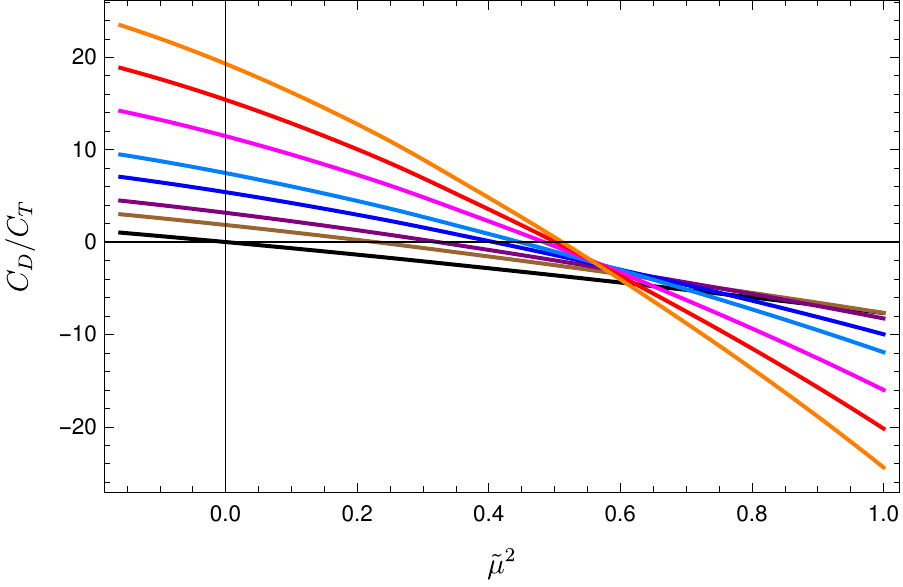}} 
\subfigure[]
{ \includegraphics[scale=0.82]{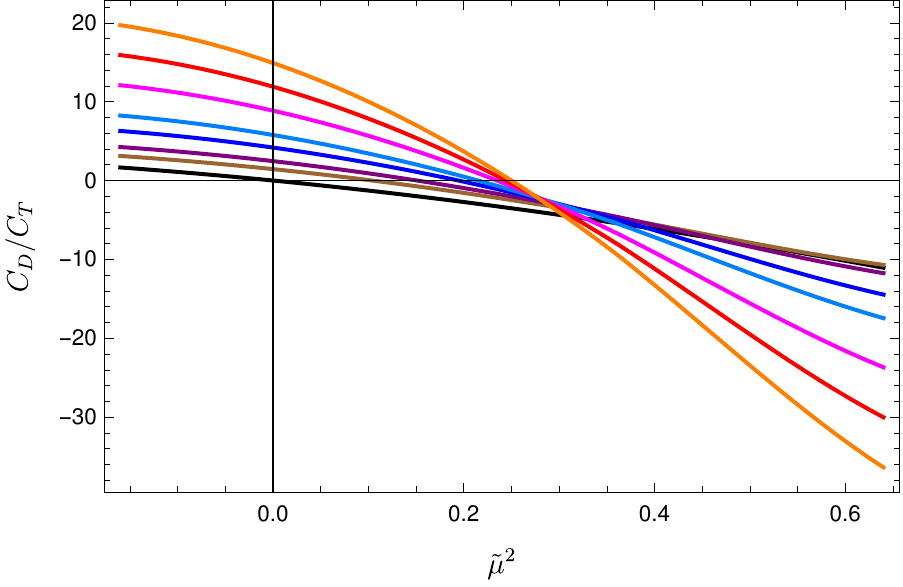}} \\
\caption{Numerical results for $C_D/C_T,$ defined in eqs.~\eqref{eq:CTCV_holography} and \eqref{eq:holographic_equation_CD}, as a function of $\tilde{\mu}^2,$ introduced in eq.~\eqref{eq:definition_mutilde}, in $d=3$ (left) and $d=4$ (right). We set $\ell_{*}/L=1.$
    Different colours correspond to different values of $n.$ The black curve corresponds to $n=1$, while the upper orange curve corresponds to $n=10$. In between, the curves go in increasing order of $n$ taking values of 
    $n=1.5,2,3,4,6,8$. Note that $\tilde{\mu}^2$ can be negative, as the chemical potential can take imaginary values.}
  \label{fig:CD3_various_n}
\end{figure}

Instead of fixing $\tilde{\mu}$, in fig.~\ref{fig:CD3_various_n} we fix $n$ and plot $C_D$ for different values of $\tilde{\mu}^2$ for $d=3$, $4$. 
Starting from the upper left side of each plot, $n$ decreases in each curve, until reaching the black curve that corresponds to $n=1,$ which, as expected, passes through the origin. We notice that $C_D$ is always a decreasing function of $\tilde{\mu}^2,$ but there is an intersection point after which the behaviour of the curves at different Rényi indices changes.

\begin{figure}[h]
    \centering
    \includegraphics[scale=1]{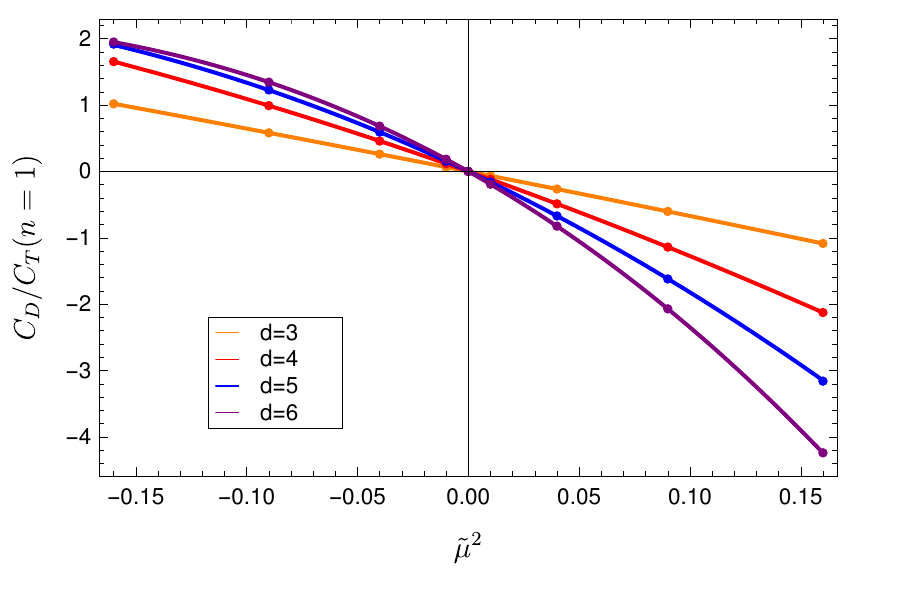}
    \caption{$C_D/C_T (n=1)$ 
    as a function of the chemical potential $\tilde{\mu}^2$ in eq.~\eqref{eq:definition_mutilde} for different spacetime dimensions.
    We set $\ell_{*}/L=1.$ 
    The dots correspond to numerical results; the solid curves correspond to the analytic expansions determined in eqs.~\eqref{eq:analyitic_expansions_CD_over_CT_d3}, \eqref{eq:analytic_result_d4_n1}, \eqref{eq:analyitic_expansions_CD_over_CT_d5} and \eqref{eq:analyitic_expansions_CD_over_CT_d6}. 
    The plot contains both real and imaginary chemical potentials such that $\tilde{\mu}^2  = \pm 0.16, \pm 0.09, \pm 0.04, \pm 0.01, 0$. }
    \label{fig:CD_fixed_n1}
\end{figure}

Finally, in fig. \ref{fig:CD_fixed_n1}, we concentrate on the case of $n=1$ and plot $C_D$ as a function of $\tilde{\mu}$ in different dimensions. The dots correspond to numerical results, while the continuous curves are the analytic expansions found in section \ref{sect-analytic}.
We observe that there is a remarkably precise agreement with the analytic expansion, which also holds for larger values of the chemical potential. All of these curves pass through the origin, since the defect disappears when $n=1$ and $\tilde{\mu}=0,$ and then, of course,  there is no effect to its deformation.

\pagebreak

\section{Discussion} \label{discussion}

In this work, we used holography to evaluate the coefficient $C_D$ appearing in the two-point function of the displacement operator  \eqref{eq:two_point_function_DD}. This coefficient completely fixes the change of charged \eqref{CREmain} and symmetry-resolved \eqref{symmetryresolvedRN}  R\'enyi entropies under small shape deformations of a flat or spherical entangling surface.

We first summarize the main results of our investigation.
Following the methods of \cite{Belin:2013uta,Bianchi:2016xvf},
an appropriate conformal transformation can be used to map the vacuum state with a slightly deformed flat entangling surface to a grand canonical ensemble on a deformed version of the manifold $S^1 \times H^{d-1}$. In that case the dual gravitational setup corresponds to a deformed black hole solution with hyperbolic horizon and a non-trivial gauge connection.
We computed the corresponding equations of motion for the deformed background and solved them both numerically and in an analytic expansion around $n=1$ and $\mu=0$.
By using holographic renormalization, we then extracted $C_D$ from the one point function of the stress-energy tensor in the deformed background.

Our numerical analysis and analytic expansions around $n=1$ and $\mu=0$ show that, for generic values of the chemical potential and of the Rényi index, $C_D$ does not generically obey the conjectured relation \eqref{eq:conj_CD}, 
originally proposed in \cite{Bianchi:2015liz}.
This is similar to what happens in the case without charge  \cite{Bianchi:2016xvf}, however, here the conjecture  is already violated at order 0 in $(n-1)$ due to the presence of the chemical potential. However, for unitary quantum field theories in the regime $\mu^2<0$ and for $n>1,$ the violation is mild for a large range of chemical potentials and replica numbers. It would be interesting to understand the reason for this.

On the other hand, we demonstrated that in the supersymmetric case, once the chemical potential is related to the number of replicas via eq.~\eqref{eq:chemical_potential_SUSY}, the 
conjecture \eqref{eq:conj_CD} holds
identically for all values of the Rényi index in various dimensions.\footnote{This claim was  proven rigorously in $d=4$ in  \cite{Bianchi:2019sxz}. In higher dimensions it is believed to be true, but it has not been proven.}
We proved this numerically in dimensions $3\leq d \leq 6,$ and analytically in an expansion around $n=1$ and $\mu=0$ (up to the order where we truncate the series) in dimensions $3\leq d \leq 7$. The holographic setup is similar in different dimensions, and we therefore view this as an indication that the conjecture holds in holography in general dimensions $d \geq 3$. This includes cases with $d>6$ where there is no dual superconformal field theory.

Once we know $C_D$, the variation of the charged Rényi entropy \eqref{CREmain}, evaluated at fixed chemical potential, is given in eqs.~\eqref{varpart}-\eqref{eq:variation_Sn_in_terms_of_CD}. An example of how to implement this variation for a deformation of a spherical entangling surface in 3 dimensions can be found in appendix \ref{app:example_3d}. Similarly, the variation of the symmetry-resolved Rényi entropy \eqref{symmetryresolvedRN}, evaluated at fixed charge, reads
\begin{equation}
   \delta S_n(q) = \frac{1}{n-1}\left(n\frac{\delta\mathcal{Z}_1(q)}{\mathcal{Z}_1(q)} - \frac{\delta\mathcal{Z}_n(q)}{\mathcal{Z}_n(q)}\right) \, ,
   \qquad \delta \mathcal{Z}_n(q) \equiv 
   -i n \int_{-i\pi/n}^{i\pi/n} \frac{d\mu}{2\pi} e^{- q n \mu} \delta Z_n(\mu) \, .
\end{equation}
To evaluate this variation, we need, in addition to the variation of $\delta \log Z_n(\mu)$ (which is fully fixed in terms of the coefficient $C_D$, see  eq.~\eqref{varpart}), also the partition function $Z_n(\mu)$ itself.
In the grand-canonical ensemble, the partition function is simply fixed in terms of the grand potential $\mathcal{G}$ in eq.~\eqref{eq:Gibbs_free_energy} by means of the identity $\mathcal{G} = - T \log Z(\mu)$.
In this way, we can extract the charged Rényi entropy and its variation for a fixed chemical potential or for a fixed charge. 

\

The investigations considered in this work open the possibility for several future directions, both from the gravity and the quantum perspective.

\

\noindent {\bf Field-theoretic outlook.}
It would be insightful to test our general results and calculate $C_D$ in explicit quantum field theories.
The computation of charged Rényi entropies has been performed in 1+1 dimensional systems, involving free scalars or fermions, both in the relativistic and in the non-relativistic scenario \cite{Belin:2013uta,Bonsignori:2019naz,Murciano:2020lqq,Murciano:2020vgh}. It would be interesting to generalize those studies to higher dimensions where deformations of the entangling surface can be performed. Recall, that in the case without a global symmetry, free theories respect the conjecture \eqref{eq:conj_CD}, as demonstrated  in specific models in $d=3,4$  \cite{Bianchi:2015liz,Dowker:2015pwa,Dowker:2015qta} and it would be interesting to check if this persists when including the effect of a global charge.  

Furthermore, we could try to construct  supersymmetric examples by tuning the chemical potential, similarly to what we did in holography in eq.~\eqref{eq:chemical_potential_SUSY}. In particular, in dimensions $d\geq 3$, we could check if the conjecture \eqref{eq:conj_CD} is satisfied due to supersymmetry.
If it is indeed the case, it is worth considering weakly-coupled field theories to confirm that the conjecture is satisfied due to supersymmetry and not because of working with free field theories.

In the present work, we used holography to prove that the conjecture \eqref{eq:conj_CD} holds for supersymmetric black holes dual to field theories in dimensions $3 \leq d \leq 7.$
The field-theoretical proof of this statement is only known in dimension $d=4$ (in addition, there are further checks in $d=3$), so one can be tempted to further generalize these results to  other dimensions, following the approach of \cite{Bianchi:2019sxz}.

\ 

\noindent {\bf Gravity outlook.}
We considered a grand-canonical ensemble at constant chemical potential. From the gravitational perspective, this is because the action \eqref{eq:EM_action} provides a well defined variational principle if we fix the dual gauge field at the boundary. Therefore, our setup  gives easy access to the charged Rényi entropies $S_n (\mu)$.
On the other hand, it might be easier to determine the symmetry-resolved Rényi entropies $S_n (q)$ by considering an ensemble where the charge is fixed.
This can be obtained in the gravitational theory with the addition of a boundary term to the action \cite{Hawking:1995ap,Chamblin:1999tk,Hartnoll:2009sz} 
\beq
\frac{\ell_*^2}{2 \ell_{\rm P}^{d-1}} \int_{\del \mathcal{M}}  \sqrt{h}  \, n^a F_{ab}   A^b \, ,
\eeq
where the indices $a,b$ run over the coordinates of the codimension-one boundary.
It would be interesting to develop this formalism further.

In the present work, we discussed shape deformations of a flat entangling surface from holography in Einstein-Maxwell gravity.
We could enlarge the investigation to Gauss-Bonnet gravity or other  higher-derivative theories \cite{Cvetic:2001bk,Anninos:2008sj,Cano:2022ord}.
In the case without charge, the conjecture \eqref{eq:conj_CD} can be satisfied up to second order around $n=1$ when the coupling $\lambda$ of the higher-derivative term is tuned to saturate the unitarity bound \cite{Bianchi:2016xvf,Chu:2016tps}. 
A natural direction would be to check if the discrepancy between $C_D$ and $C_D^{\rm conj}$, which we observed in the case with non-vanishing chemical potential, can be improved around $n=1$  for specific values of the Gauss-Bonnet coupling. We would also like to check  whether the conjecture in the supersymmetric case holds in Gauss-Bonnet gravity.

Finally, as briefly discussed in \cite{Belin:2013uta}, another possible generalization of the Rényi entropies involves the case of
a spherical entangling surface, where the states are labelled by their angular momenta instead of the charge.
The dual gravitational configuration would be a spinning hyperbolic black hole, with an associated rotating Rényi entropy.
We could also study shape deformations of spherical defects in this context.

\

\noindent {\bf Experimental outlook.}
While Rényi entropies may naively appear  as  abstract quantities, they have some recent concrete applications.
For example, the second Rényi entropy was measured in an experiment involving ultra-cold bosonic atoms in optical lattices \cite{islam2015measuring}.
Essentially, the idea is the following. 
One prepares two identical copies of a state composed by $N$ particles and interferes them using a double well potential.
By measuring the probability of finding an even/odd number of the particles in one copy or the other after the interference, one determines the overlap between the two systems, which is related  to the second Rényi entropy.
A generalization of this procedure  including an Aharonov-Bohm flux was discussed in \cite{Goldstein:2017bua}.
The proposal is to realize the same experimental set-up, but now restricting to sectors of fixed charge and averaging between them.

The measure of charged Rényi entropies allows to distinguish symmetry-protected topological states from other phases of matter, thanks to the degeneracies that are present in their entanglement spectrum \cite{Azses:2020tdz}.
As an example, the authors of  \cite{Azses:2020tdz} demonstrated how to implement a protocol on the IBM quantum computer to identify the symmetry-protected nature of the ground state of a one-dimensional cluster Ising Hamiltonian.
By making two copies of the system and performing certain swap operations between them, it is possible to extract the second Rényi entropy and its restriction to the charge sectors.
It would be interesting to perform similar simulations in higher dimensions where the entangling surface can be deformed, which might bring the fascinating possibility of being able to test quantum field theory and/or gravitational results in quantum simulations.

\section*{Acknowledgements}

We gratefully acknowledge discussions  with Dionysios  Anninos, Tarek Anous, Igal Arav, Ramy Brustein, Lorenzo Di Pietro, Zohar Komargodski, Michael Lublinsky, Marco Meineri, Tatsuma Nishioka,  Eran Sela and Chiara Toldo. The work of SC and SB is supported by the Israel Science Foundation (grant No. 1417/21) and by the German Research Foundation through a German-Israeli Project Cooperation (DIP) grant “Holography and the Swampland”. 
SB is supported by the Kreitmann School of Advanced Graduate Studies and by the Azrieli Foundation.
SC acknowledges the support of Carole and Marcus Weinstein through the BGU Presidential Faculty Recruitment Fund. The research of LB is funded through the MIUR program for young researchers “Rita Levi Montalcini”.
The work of DAG is funded by the Royal Society under the grant ``The Resonances of a de Sitter Universe'' and the ERC Consolidator Grant N. 681908, ``Quantum black holes: A microscopic window into the microstructure of gravity''.
DAG is also funded by a UKRI Stephen Hawking Fellowship.

\appendix

\section{Explicit example in 3 dimensions: deformation of a circle}
\label{app:example_3d}

\begin{figure}[h]
\centering
\subfigure[]
{ \includegraphics[scale=0.73]{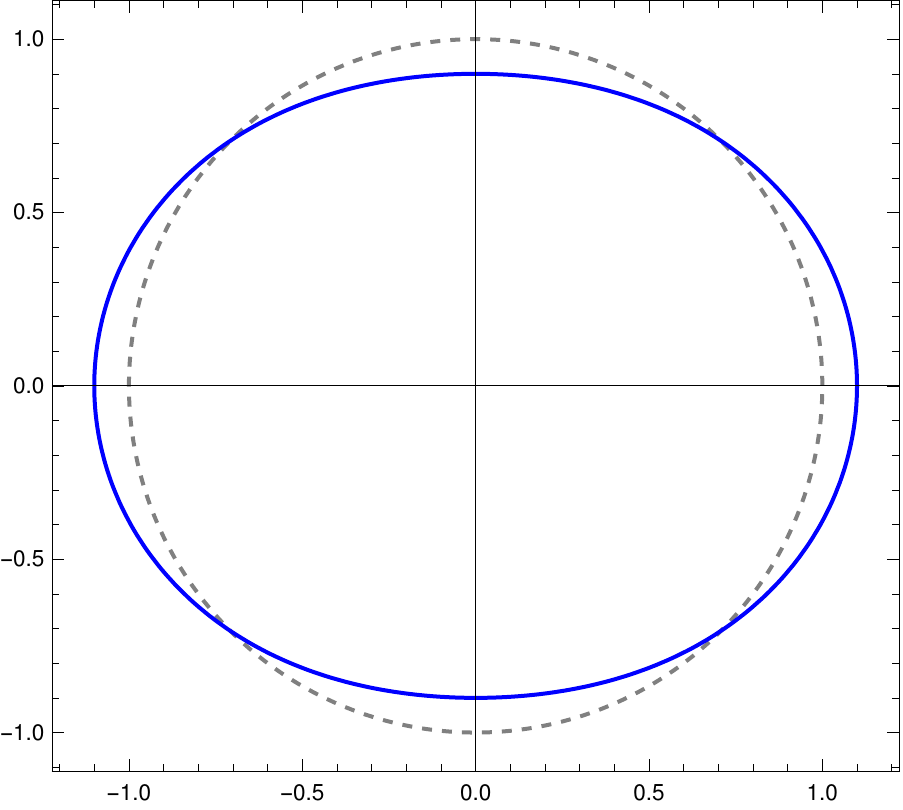}} 
\subfigure[]
{ \includegraphics[scale=0.68]{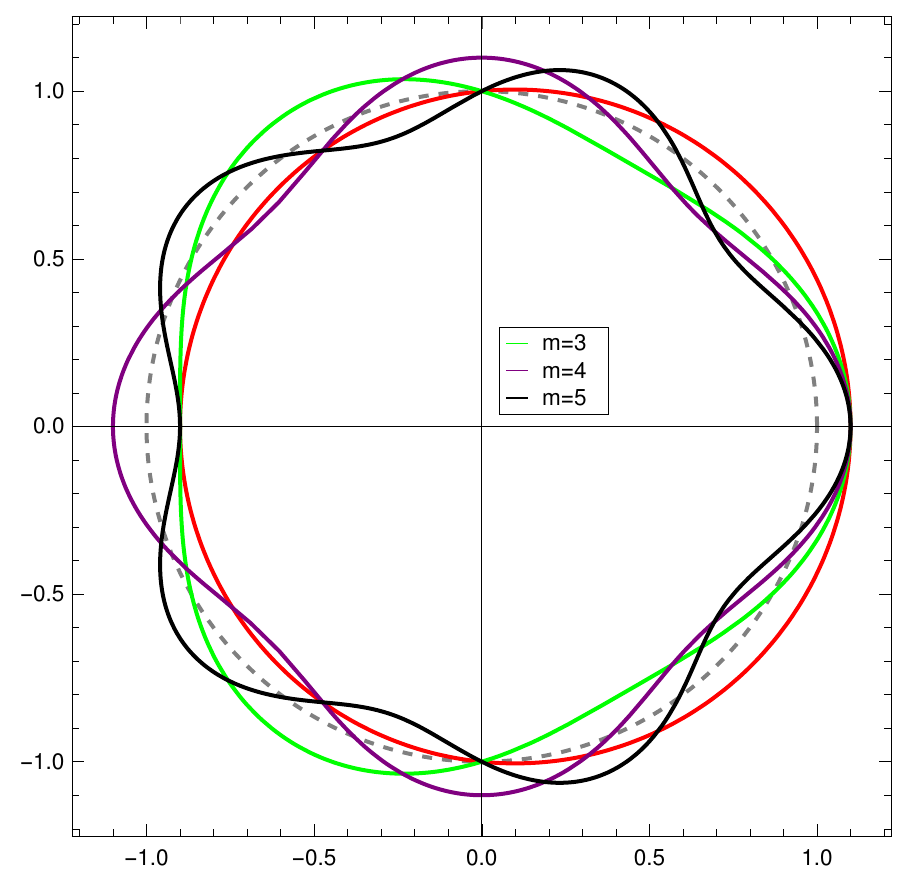}} \\
\caption{(a) Deformation of a circle (dashed gray) into an ellipse (blue).
(b) Deformations from the circle (dashed gray) to other curves, using a deformation $\delta X^r = R \varepsilon  \cos (m \theta).$ Here we fixed $\varepsilon=0.1$ and $R=1.$
The red curve corresponds to $m=1.$ Since it is simply a shift of the original circle, it does not contribute to the deformation of the partition function in eq.~\eqref{eq:example_CD_d3}.}
\label{fig:example_3d}
\end{figure}

In this appendix, we consider the shape deformation of a circular entangling surface in a three-dimensional CFT to provide a simple example where $C_D$ can be explicitly related  to the variation of the R\'enyi entropy. Let us consider a timeslice of three-dimensional spacetime parametrized by polar coordinates $(r,\theta)$. The entangling surface lies at $r=R$ and we can consider a $\theta$-dependent deformation in the radial direction
\beq
\delta X^r= R \, \varepsilon f(\theta) \, ,
\label{eq:variation_example_3d}
\eeq
where $\varepsilon$ is a small dimensionless parameter and $f(\theta)$ is a generic periodic function of $\theta$. Comparing with equation \eqref{deformation} one notices that this is not the most general deformation since we have two orthogonal directions, the radial and the time direction. Nevertheless, for concreteness, we focus here on the most natural shape deformation for an entangling surface, \ie the one that does not extend in the time direction. Furthermore, we expand the deformation $f(\theta)$ as
\begin{equation}
f(\theta)=\sum_m f_m \cos \le m\theta \ri \, ,
\end{equation}
with $m \in \mathbb{N}.$ 
For instance, the case where the circle is deformed into an ellipse corresponds to $f_m=\delta_{m,2}$. The changes of the shape for various choices of $m$ are depicted in fig.~\ref{fig:example_3d}.

Using the identity \eqref{varpart}, we obtain the variation of the partition function
\beq\label{eq:3dvariation}
\frac{d^2 \log Z_n (\mu)}{d \varepsilon^2}\Big|_{\varepsilon=0}  = R^4 \sum_{m,m'} f_m f_{m'} \int_0^{2 \pi} d\theta \int_0^{2 \pi} d\theta' \, 
\cos (m \theta) \cos (m' \theta')\langle D^r (\theta) D^r(\theta')\rangle \, ,
\eeq
where $D^r$ is the radial component of the displacement operator and the overall factor of $R^4$ comes from the normalization in eq.~\eqref{eq:variation_example_3d} and the integration along the dimensionful coordinate $w=R \theta$. The two-point function of $D^r$ can be easily obtained with the appropriate conformal transformation in \eqref{eq:two_point_function_DD},
\beq
\langle D^r(\theta) D^r(0) \rangle_{n,\mu} = \frac{C_D (n,\mu)}{4 R^4 \le 1- \cos \theta \ri^2 } \, .
\eeq
Substituting this into eq.~\eqref{eq:3dvariation} one can easily check that the integral vanishes unless $m=m'$ and we get the universal contribution
\beq
\begin{aligned}
\frac{d^2 \log Z_n (\mu)}{d \varepsilon^2}\Big|_{\varepsilon=0} & =\sum_m f_m^2 C_D (n,\mu) \int_0^{2 \pi} d\theta \int_0^{2 \pi} d\theta' \, 
\frac{\cos (m \theta) \cos (m \theta')}{4 \le 1- \cos \le \theta - \theta' \ri \ri^2} = \\
& =\sum_{m} f_m^2 C_D (n,\mu) \, \frac{\pi}{4} \int_0^{2 \pi} d\theta \, \frac{\cos \le m \theta \ri}{\le 1 - \cos \theta \ri^2} \\
& = 
\frac{\pi^2}{6} C_D (n,\mu) \sum_m f_m^2 m (m^2 -1) \, .
\end{aligned}
\label{eq:example_CD_d3}
\eeq
In going from the first to the second line, we made use of the prosthaphaeresis formula $ \cos (m \theta) \cos (m \theta') = \frac{1}{2} \le \cos (m (\theta+ \theta')) - \cos (m (\theta-\theta')) \ri .$
The former term vanishes because it reduces to an odd integral along an even interval.
The latter term is evaluated explicitly.
The expression is formally divergent, but its finite part is universal, because in odd spacetime dimensions there is no logarithmic divergence.
After subtracting the divergence, we obtain the contribution reported above.\footnote{We use a cutoff regularization: introducing the infinitesimal parameter $\delta,$ we perform the integration along the region $\theta \in [\delta, 2 \pi - \delta]$ and then we perform a Laurent-expansion around $\delta=0$ to isolate the divergence. Alternatively, one can use dimensional regularization, which is blind to power-law divergences.} Equation \eqref{eq:example_CD_d3} shows that $C_D$ precisely accounts for the second-order shape deformation of the entangling surface. For instance for the elliptic deformation
\beq
\frac{d^2 \log Z_n (\mu)}{d \varepsilon^2}\Big|^{\text{ellipse}}_{\varepsilon=0} =\frac{\pi^2}{3}  C_D(n, \mu) \, .
\eeq
In order to get the variation of the Rényi entropy $\delta S_n(\mu),$ one has to take the value of $C_D$ computed at the desired Rényi index and chemical potential looking at the $d=3$ analysis in section \ref{sect-numerical}. 
Then one plugs it inside eq.~\eqref{eq:variation_Sn} using the result \eqref{eq:example_CD_d3}.

\section{On-shell action for charged hyperbolic black holes}
\label{app:on_shell_action}

In this appendix, we evaluate the regularized on-shell action $I_0$ for hyperbolic charged black holes in Einstein-Maxwell gravity in general spacetime dimensions. This computes the grand-canonical partition function $Z(\mu)$ according to
\begin{equation}
    I_{0} = -\log Z (\mu) \, .
\end{equation}

We start from the gravitational action $I_{\rm EM}$ in eq.~\eqref{eq:EM_action},
\beq
I_{\rm EM} = - \frac{1}{2 \ell_P^{d-1}} 
\int_{\mathcal{M}} d^{d+1} x \, \sqrt{g} \le \mathcal{R} + \frac{d(d-1)}{L^2} - \frac{\ell_*^2}{4} F_{\mu\nu} F^{\mu\nu}  \ri \, ,
\label{eq:EM_action_app}
\eeq
 supplemented by the usual boundary Gibbons-Hawking-York (GHY) term,
\beq
I_{\rm GHY} = - \frac{1}{\ell_{\rm P}^{d-1}} \int_{\mathcal{B}} d^d x \, \sqrt{h} \, K \, ,    
\eeq
where $h_{ab}$ denotes the induced metric on the boundary $\mathcal{B}=\del \mathcal{M}$ of the bulk manifold, and $K$ is the trace of the extrinsic curvature.

As we will see, the bare on-shell action $I_0+I_{\rm GHY}$ is divergent due to contributions close to the AdS boundary. We will regulate the on-shell action by the standard holographic method of adding local boundary counterterms to the action \cite{Emparan:1999pm,Chamblin:1999hg}.\footnote{Earlier in the literature, these divergences were regulated by background subtraction, see \eg \cite{Chamblin:1999tk,Cai:2004pz}.}
The counterterm action is then given by
\beq
\begin{aligned}
I_{\rm ct} = \frac{1}{\ell_{\rm P}^{d-1}} \int_{ \mathcal{B}} d^d x \, \sqrt{h}  \, & \left[  \frac{d-1}{L} + \frac{L}{2(d-2)} \mathcal{R}_{\mathcal{B}}  \right. \\
& \left.
+ \frac{L^3}{2(d-4)(d-2)^2} \le \mathcal{R}_{ab} \mathcal{R}^{ab} - \frac{d}{4(d-1)} \mathcal{R}^2_{\mathcal{B}} \ri + \dots  \right] \, ,
\end{aligned}
\label{eq:counterterm}
\eeq
where $\mathcal{R}_{ab}$ is the Ricci tensor for the boundary metric and $\mathcal{R}_{\mathcal{B}}$ the corresponding Ricci scalar.
The ellipsis denote additional terms which need to be included to subtract the divergences when $d > 7.$
The renormalized on-shell action is then given by the sum of all the previous terms, \ie
\beq
I_{0} = I_{\rm EM} + I_{\rm GHY} + I_{\rm ct} \, .
\eeq

Next, we provide details of the explicit calculation in $d=3.$
We parametrize the metric in terms of $(\tau,r,\rho,y)$ coordinates as
\beq
ds^2 = \frac{L^2}{R^2} \, G(r) d\tau^2 + \frac{dr^2}{G(r)} + \frac{r^2}{\rho^2} \le d\rho^2 + dy^2 \ri \, ,
\eeq
where $G(r)$ is the blackening factor in eq.~\eqref{eq:blackening_factor_Einstein_Maxwell}.
By plugging the solutions for the metric and the gauge connection \eqref{eq:zeroth_order_solution_gauge} in the bulk action \eqref{eq:EM_action_app}, we find
\beq
\begin{aligned}
I_{\rm EM} & = \frac{1}{\ell_{\rm P}^2} 
\int_{\mathcal{M}} d^d x \,   \frac{3r^4 - L^2 Q^2}{LR \, r^2 \rho^2} =
 \frac{1}{LR \, \ell_{\rm P}^2}  \frac{V_{\Sigma}}{T}   \int_0^{r_{\rm max}} dr \,    \frac{3r^4 - L^2 Q^2}{r^2} = \\
 & =  \frac{1}{LR \, \ell_{\rm P}^2}  \frac{V_{\Sigma}}{T} 
 \left[ r_{\rm max}^3  - \le L^2 Q^2 + r_h^4 \ri  \right] + \mathcal{O} (r_{\rm max}^{-1}) \, .
 \end{aligned} 
\eeq
In the first line we performed explicitly the integration along the hyperbolic space $H^2,$ giving the regulated dimensionless volume $V_{\Sigma},$ and we integrated along the Euclidean time direction, which brings a factor of $T^{-1}$ due to the periodicity of the thermal circle.
In the second line we performed the integral along the radial coordinate and expanded the solution around a UV cutoff $r=r_{\rm max}.$ Indeed, the bulk action is divergent as $r_{\rm max} \to \infty$.

In order to evaluate the GHY contribution and the counterterm, we need to specify the induced metric data on the boundary identified by the surface at constant $r=r_{\rm max}.$
The induced metric and the outgoing normal one-form to this surface are given by
\beq
ds^2_{\rm ind} = \frac{L^2}{R^2} \, g(r_{\rm max}) d\tau^2 + \frac{r_{\rm max}^2}{\rho^2} \le d\rho^2 + dy^2 \ri \, , \qquad
\mathbf{n} = \frac{d\tau}{\sqrt{G(r)}} \, .
\eeq
These data are sufficient to determine the extrinsic curvature, the induced metric determinant and the Ricci tensor on the boundary.
By direct computation, we find
\beq
I_{\rm GHY} = \frac{1}{LR \, \ell_{\rm P}^2}  \frac{V_{\Sigma}}{T} 
 \left[ -3 r_{\rm max}^3 + 2 L^2 r_{\rm max} + 3 L^2 M \right] + \mathcal{O} (r_{\rm max}^{-1}) \, ,
\eeq
\beq
I_{\rm ct} = \frac{1}{LR \, \ell_{\rm P}^2}  \frac{V_{\Sigma}}{T} 
 \left[ 2 r_{\rm max}^3 - 2 L^2 r_{\rm max} -  L^2 M \right] + \mathcal{O} (r_{\rm max}^{-1}) \, .
\eeq
We point out that the contribution in the second line in eq.~\eqref{eq:counterterm} does not modify the result in $d=3,$ but it plays an important role to modify the constant term in $d=4,$ and to cancel the divergences in $d \geq 5.$
Summing all the terms entering the renormalized action and using eqs.~\eqref{eq:relation_mass_charge} and \eqref{eq:chemical_potential_and_charge}, we see that all the divergences cancel and we obtain
\beq
I_0 = - \frac{V_{\Sigma} r_h}{\ell_{\rm P}^2}  \frac{1}{T} 
\left[ 1 + \frac{1}{4} \le \frac{\mu \ell_*}{2 \pi L} \ri + \frac{r_h^2}{L^2} \right] \, ,
\eeq
which is finite.
Now it is easy to obtain the grand potential, since 
\beq
\mathcal{G} = -T \log Z(\mu) = T I_0 \, . 
\eeq
One can perform the same computation in any dimension $d \leq 7$. The general result reads
\beq
\mathcal{G} = - \frac{V_{\Sigma} r_h^{d-2}}{2 \ell_{\mathrm{P}}^{d-1}} 
\left[   1+ \frac{d-2}{2(d-1)} \le \frac{\mu \ell_*}{2 \pi L} \ri^2  
+ \frac{r_h^2}{L^2}
\right] \, ,
\eeq
which is precisely eq.~\eqref{eq:Gibbs_free_energy}.

\section{Details of the holographic renormalization} \label{app:details_holo_ren}

In order to determine the precise dictionary between $\beta_n (\mu)$ introduced in the expansions \eqref{eq:asymptotic_expansion_k} and $C_D,$  we apply the holographic renormalization procedure explained in section \ref{sect-holo_renormalization}.
In this appendix we add further details of the derivation.
The first step is to put the metric in the FG form \eqref{eq:metric_FG_expansion}.
This is achieved by performing a double change of variables
\beq
r = \frac{L^2}{\tilde{z}} \, , \qquad
\tilde{z} = z \le 1 + c_1 \frac{z}{L} + c_2 \frac{z^2}{L^2} + \dots + c_d \frac{z^d}{L^d} + \dots  \ri \, ,
\eeq
where the coefficients $c_i$ are determined order by order in the expansion by imposing that the radial part of the metric is
\beq
\frac{dr^2}{G(r)} =
\frac{L^4 dz^2 \le \del \tilde{z}/\del z \ri^2}{\tilde{z}^4 G(r(z))} 
= \frac{L^2}{z^2} \, dz^2 \le 1 + \mathcal{O}(z^{d+1}) \ri \, .
\label{eq:change_blackening_factor_FG_expansion}
\eeq 
The resulting transformations in various dimensions are given by
\beq
\begin{aligned}
&  d=3 \, , \quad \tilde{z} = z \left[ 1 - \frac{z^2}{4 L^2} - \frac{1}{6} x_n \le x_n^2 -1 + \le \frac{\mu \ell_*}{2 \pi L} \ri^2  \ri \frac{z^3}{L^3} + \mathcal{O} (z^4) \right] \, , & \\
& d=4 \, , \quad \tilde{z} = z \left[ 1 - \frac{z^2}{4 L^2} -
\frac{1}{48} \le 6 x^2 \le x_n^2-1 \ri  +  2 x_n^2  \le \frac{\mu \ell_*}{2 \pi L} \ri^2  -3 \ri  \frac{z^4}{L^4} + \mathcal{O} (z^5) \right] \, , & \\
& d=5 \, , \quad \tilde{z} = z \left[ 1 - \frac{z^2}{4 L^2} + \frac{z^4}{16 L^4} - \frac{1}{20} x_n^3 \le 2 \le x_n^2 -1 \ri +3 \le \frac{\mu \ell_*}{2 \pi L} \ri^2   \ri  \frac{z^5}{L^5} + \mathcal{O} (z^6) \right] \, , & \\
& d=6 \, , \quad \tilde{z} = z \left[ 1 - \frac{z^2}{4 L^2} + \frac{z^4}{16 L^4} -
\frac{1}{960} \le 80 x_n^4 \le x_n^2 -1 \ri + 128 x_n^4 \le \frac{\mu \ell_*}{2 \pi L} \ri^2   +15 \ri \frac{z^6}{L^6} + \mathcal{O} (z^7) \right] \, , & \\
& d=7 \, , \quad \tilde{z} = z \left[ 1 - \frac{z^2}{4 L^2} + \frac{z^4}{16 L^4} - \frac{z^6}{64 L^6}
- \frac{1}{42} x_n^5 \le 3 \le x_n^2 -1 \ri + 5  \le \frac{\mu \ell_*}{2 \pi L} \ri^2   \ri  \frac{z^7}{L^7}
 + \mathcal{O} (z^8) \right] \, . & \\
\end{aligned}
\label{eq:change_variables_FG_4dim}
\eeq
We notice that the chemical potential enters explicitly these transformations, and they reduce to the expressions listed in appendix B of \cite{Bianchi:2016xvf} when $\mu=0.$
The dependence on both the chemical potential and the number $n$ of replicas (via the quantity $x_n$) starts only at order $z^d.$
Therefore, all the lower-order terms $\mathfrak{h}^{(m)}$ in the expansion of the boundary metric \eqref{eq:metric_FG_expansion} will not depend on either $\mu$ and $n,$ and the same reasoning applies to the functional $\mathcal{X}_{\mu\nu}$ defined in eq.~\eqref{eq:holographic_stress_tensor} as well.

Now, we apply the changes of variables \eqref{eq:change_variables_FG_4dim} to the metric \eqref{eq:bulk_metric_ansatz}. We find
\beq
\begin{aligned}
\mathfrak{h}_{\mu\nu} dx^{\mu} dx^{\nu}  = & \, G(r(z)) \frac{R^2}{L^2} d\tau^2 + L^2 \le \frac{z}{\tilde{z}} \ri^2 \frac{d \rho^2}{\rho^2} +
\left[\frac{4 L^2}{d-2} \le \frac{z}{\tilde{z}} \ri^2 v \le r(z) \ri \right] \le x_a \, \del_i K^a  \ri \frac{d\rho}{\rho} \, dy^i \\
& + \frac{L^2}{\rho^2} \left[  \le \frac{z}{\tilde{z}} \ri^2 \delta_{ij} +  2 \le \frac{z}{\tilde{z}} \ri^2 k(r(z)) \, \tilde{K}^a_{ij} x_a   \right] dy^i dy^j \, .
\end{aligned}
\label{eq:boundary_metric_holo_ren}
\eeq
By identifying the order $d$ terms in the previous expansion, we determine the coefficients entering the one-point function of the stress tensor defined in eq.~\eqref{eq:CFToutputkn}. 
The solutions for the lower-order terms are not needed for the holographic determination of $C_D.$
The interested reader can find their expressions in \cite{deHaro:2000vlm,Bianchi:2016xvf}.\footnote{Since the lower-order terms do not depend on the chemical potential, the same results as in the case without charge hold.}

\section{Solutions to the differential equations in higher dimensions} \label{app:diff_eq_higherd}

In this appendix, we provide further details on the calculation of the analytic and numerical solutions to the differential equations \eqref{eq:differential_equations_G} in $d=5,6,7.$
The techniques used to derive these results are the same outlined in sections \ref{sect-analytic} and \ref{sect-numerical}.

\subsection{Analytic expansions}

In higher dimensions, the asymptotic expansions of the solution to the differential equation \eqref{eq:differential_equations_G} read
\beq
\begin{aligned}
& d=5 \, , \qquad  k(r)= v(r) = 1 - \frac{L^2}{2 r^2} - \frac{L^4}{8 r^4} + \frac{L^5}{r^5} \beta_n (\mu) + \mathcal{O}(r^{-6}) \, , & \\
& d=6 \, , \qquad  k(r)= v(r) = 1 - \frac{L^2}{2 r^2} - \frac{L^4}{8 r^4}  + \frac{L^6}{r^6} \beta_n (\mu) + \mathcal{O}(r^{-7}) \, , &  \\
& d=7 \, , \qquad  k(r)= v(r) = 1 - \frac{L^2}{2 r^2} - \frac{L^4}{8 r^4}  - \frac{L^6}{16 r^6} + \frac{L^7}{r^7} \beta_n (\mu) + \mathcal{O}(r^{-8}) \, . & 
\end{aligned}
\label{eq:asymptotic_expansion_k_higher_d}
\eeq
We can obtain an analytic expression around $n=1$ and $\mu=0$ for the undetermined coefficient $\beta_n$ by considering the following analytic expansion of the solution to the differential equation, \ie
\beq
k(\tilde{r})  =  \sum_{A,B} k_{AB} (\tilde{r}) (n-1)^A \mu^B  \, .
\label{eq:analytic_expansion_k}
\eeq
The functions $k_{AB}$ at order 0 in the chemical potential and up to first in the Rényi index are \cite{Bianchi:2016xvf}
\beq
k_{00} (\tilde{r}) = \sqrt{1- \tilde{r}^2} \, , \qquad
k_{10} (\tilde{r}) = \frac{(d-1)(\tilde{r}^2-1) \tilde{r}^d \, {}_2F_1(1, \frac{d}{2}, \frac{d+2}{2},\tilde{r}^2) + d(\tilde{r}^d - \tilde{r}^2)}{(d-1)d \sqrt{1-\tilde{r}^2}} \, ,
\eeq
while the expression for $k_{20} (\tilde{r})$ is cumbersome and we will not report it explicitly.

We series-expand the coefficient $\beta_n (\mu)$ entering the asymptotic expansions according to eq.~\eqref{eq:analytic_expansion_beta}.
In various dimensions, it turns out that all the terms entering $\beta_n (\mu) $ linear in the chemical potential vanish.
More generally, we observe that
\beq
\forall m,n \in \mathbb{N} \qquad
\beta_{m, 2n+1} = 0 \, .
\eeq
At order 0 in the chemical potential, the lowest orders in the expansion around $n=1$ have a simple closed form \cite{Bianchi:2016xvf}
\beq
\beta_{00} = \begin{cases}
 - \frac{\Gamma (\frac{d-1}{2})}{2 \sqrt{\pi} \, \Gamma (\frac{d}{2}+1)}  & \mathrm{if} \,\,\, d \in 2 \mathbb{N} \\
 0 & \mathrm{if} \,\,\, d \in 2 \mathbb{N}+1  \\
\end{cases}
\label{eq:beta00_uncharged}
\eeq
\beq
\beta_{10} = \frac{1}{d(d-1)} \, , \qquad
\beta_{20} = - \frac{4d^3 -8 d^2 +d +2}{2d^2 (d-1)^3} \, .
\label{eq:beta10_beta20_uncharged}
\eeq
It is difficult to determine generic higher-order terms in a general number of  dimensions, but it turns out that we can work out the higher-order terms in $\mu$ at exactly $n=1.$
They correspond to the coefficients of kind $\beta_{m0}$ with $m \in \mathbb{N}.$
We list the results:
\beq
\begin{aligned}
& d=3 \, , \qquad  \beta_{02} = - \frac{5}{36} \, , \qquad
\beta_{04} = \frac{16320 \log 2 - 11417}{25920} \, , & \\
& d=4 \, , \qquad \beta_{02} = - \frac{7}{36} \, , \qquad
\beta_{04} = - \frac{815}{7776} \, ,  &  \\
& d=5 \, , \qquad  \beta_{02}= - \frac{81}{400} \, , \qquad
\beta_{04} = \frac{118863360  \log 2 - 107796391}{98560000}  \, , & \\
& d=6 \, , \qquad \beta_{02} = - \frac{44}{225} \, ,  \qquad
\beta_{04} = - \frac{730484}{1771875}  \, ,  &  \\
& d=7 \, , \qquad  \beta_{02} = - \frac{325}{1764} \, ,  \qquad 
\beta_{04} =  \frac{5(21907005120 \log 2 -25207810051)}{90774395712} \, . & 
\end{aligned}
\eeq
In dimensions $d=3,4$ the previous expressions are used to find eqs.~\eqref{eq:analyitic_expansions_CD_over_CT_d3_n1} and \eqref{eq:analytic_result_d4_n1}.
In higher dimensions, they determine the expansions of $C_D$ and $C_D^{\rm conj}$ at $n=1$ in terms of the rescaled chemical potential \eqref{eq:definition_mutilde}. We list the main results:
\begin{itemize}
    \item Case $d=5$
\beq
\frac{C_D}{C_T}\Big|_{n=1} = - \frac{8}{5} \pi^2 \tilde{\mu}^2 
+ \frac{1857240 \log 2 - 2800969}{616000} \pi^2 \tilde{\mu}^4 + \mathcal{O} (\tilde{\mu}^6) \, ,
\label{eq:analyitic_expansions_CD_over_CT_d5}
\eeq
\beq
\frac{C_D^{\rm conj}}{C_T}\Big|_{n=1} = - \frac{7}{4} \pi^2 \tilde{\mu}^2 
- \frac{1485}{512} \pi^2 \tilde{\mu}^4 + \mathcal{O} (\tilde{\mu}^6) \, .
\eeq
\item Case $d=6$
\beq
\frac{C_D}{C_T}\Big|_{n=1} = - \frac{40}{21} \pi^2 \tilde{\mu}^2 
- \frac{750496}{165375} \pi^2 \tilde{\mu}^4 + \mathcal{O} (\tilde{\mu}^6) \, ,
\label{eq:analyitic_expansions_CD_over_CT_d6}
\eeq
\beq
\frac{C_D^{\rm conj}}{C_T}\Big|_{n=1} = - \frac{72}{35} \pi^2 \tilde{\mu}^2 
- \frac{22784}{4375} \pi^2 \tilde{\mu}^4 + \mathcal{O} (\tilde{\mu}^6) \, .
\eeq
\item Case $d=7$
\beq
\frac{C_D}{C_T}\Big|_{n=1} = - \frac{15}{7} \pi^2 \tilde{\mu}^2 
+ \frac{25 \le 912791880 \log 2 - 1846809649 \ri}{4322590272}  \pi^2 \tilde{\mu}^4 + \mathcal{O} (\tilde{\mu}^6) \, ,
\eeq
\beq
\frac{C_D^{\rm conj}}{C_T}\Big|_{n=1} = - \frac{55}{24} \pi^2 \tilde{\mu}^2 
- \frac{81875}{10368} \pi^2 \tilde{\mu}^4 + \mathcal{O} (\tilde{\mu}^6) \, .
\eeq
\end{itemize}
It is clear that the value of $C_D$ and $C_D^{\rm conj}$ differ at $n=1$ in all dimensions, starting at second order in the chemical potential.
This extends the results discussed in section \ref{sect-analytic} for $d=3,4.$

In the supersymmetric scenario, the chemical potential is related to the number of replicas via eq.~\eqref{eq:chemical_potential_SUSY} and therefore we only need to specify the order in the expansion around $n=1.$
It turns out that we can investigate more easily the higher-order terms compared to dimensions $d=3,4.$
We find
\beq
\begin{aligned}
 d=5 \, , \qquad  \beta_n = & \frac{1}{20} (n-1) - \frac{7}{80} (n-1)^2 
+ \frac{43}{320} (n-1)^3 + \mathcal{O} (n-1)^4 \, , & \\
 d=6 \, , \qquad  \beta_n = & - \frac{1}{16} + \frac{1}{30} (n-1) - \frac{3}{50} (n-1)^2 + \frac{71}{750} (n-1)^3
 - \frac{173}{1250} (n-1)^4 +  \mathcal{O} (n-1)^5 \, , & \\
 d=7 \, , \qquad  \beta_n = & \frac{1}{42} (n-1) - \frac{11}{252} (n-1)^2  + \frac{53}{576} (n-1)^3 - \frac{473}{4536} (n-1)^4
+ \mathcal{O} (n-1)^5 \, , & \\
\end{aligned}
\eeq
which imply
\begin{itemize}
    \item Case $d=5$
    \beq
    \frac{C_D}{C_T} = \frac{C_D^{\rm conj}}{C_T} =  \, \frac{\pi^2}{3} (n-1) - \frac{\pi^2}{4} (n-1)^2
+ \frac{5 \pi^2}{16} (n-1)^3 + \mathcal{O}(n-1)^4 \, .
    \eeq
    \item Case $d=6$
    \beq
    \frac{C_D}{C_T} = \frac{C_D^{\rm conj}}{C_T} =  \, \frac{2 \pi^2}{7} (n-1) - \frac{8 \pi^2}{35} (n-1)^2 
 + \frac{52 \pi^2}{175} (n-1)^3
  - \frac{328 \pi^2}{875} (n-1)^4
+ \mathcal{O}(n-1)^5 \, .
    \eeq
    \item Case $d=7$
    \beq
    \frac{C_D}{C_T} = \frac{C_D^{\rm conj}}{C_T} =  \, \frac{\pi^2}{4} (n-1) - \frac{5 \pi^2}{24} (n-1)^2 
 + \frac{5 \pi^2}{18} (n-1)^3 - \frac{155 \pi^2}{432} (n-1)^4
+ \mathcal{O}(n-1)^5 \, .
    \eeq
\end{itemize}
As can be observed, in the supersymmetric case, we find a match between $C_D$ and $C_D^{\rm conj}$ at all the orders considered around $n=1.$

\subsection{Numerical results}

\begin{figure}[H]
\centering
\subfigure[]
{ \includegraphics[scale=0.82]{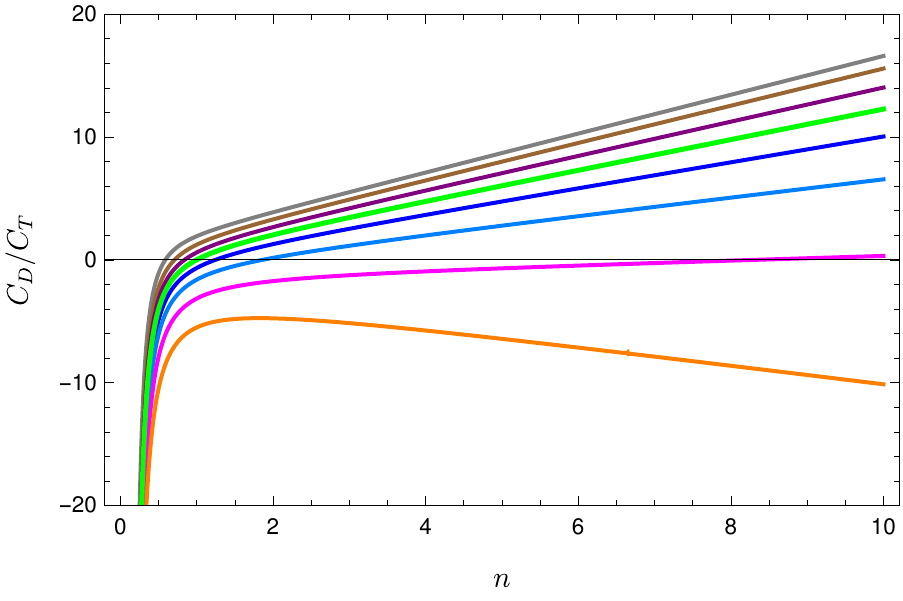}} 
\subfigure[]
{ \includegraphics[scale=0.82]{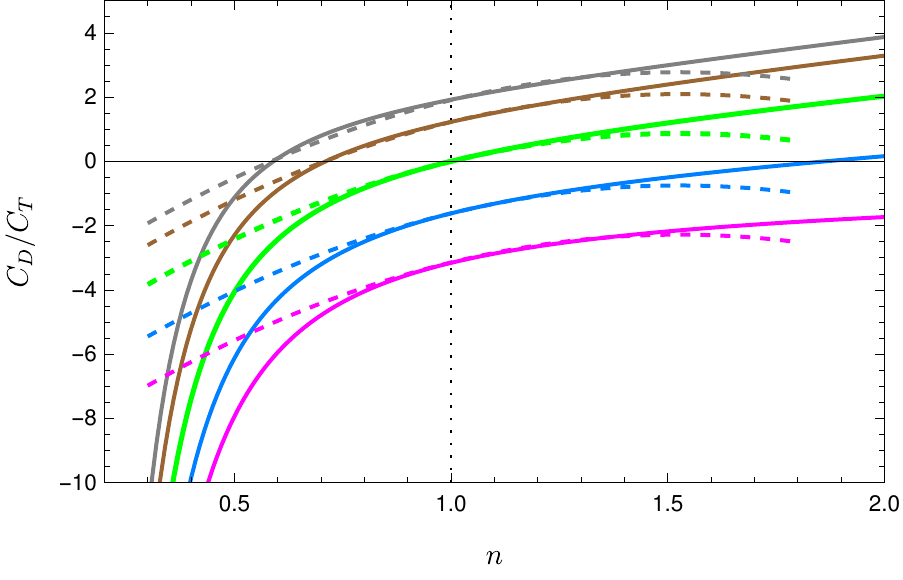}} \\
\caption{Plots of $C_D/C_T$, defined in eqs.~\eqref{eq:CTCV_holography} and \eqref{eq:holographic_equation_CD}, as a function of $n$ in $d=5.$ The colours correspond to different fixed values of the chemical potential $\tilde{\mu}$ in eq.~\eqref{eq:definition_mutilde} with $\ell_{*}/L=1.$ 
(a) The different curves correspond to $\tilde{\mu}^2 =\pm 0.04, \pm 0.09, \pm 0.16, 0.25$, with $\tilde{\mu}^2$ increasing from top to bottom. We highlight in green the curve corresponding to $\tilde{\mu}=0$. Note that $\tilde{\mu}^2$ can be negative, as the chemical potential can take imaginary values.
 (b) Zoom of the same plot around the region close to $n=1$. The dashed lines correspond to the analytic expansions in eq.~\eqref{eq:analyitic_expansions_CD_over_CT_d5}. Curves with the same colours correspond to the same chemical potential.}
\label{fig:CD5}
\end{figure}

\begin{figure}[H]
\centering
\subfigure[]
{ \includegraphics[scale=0.82]{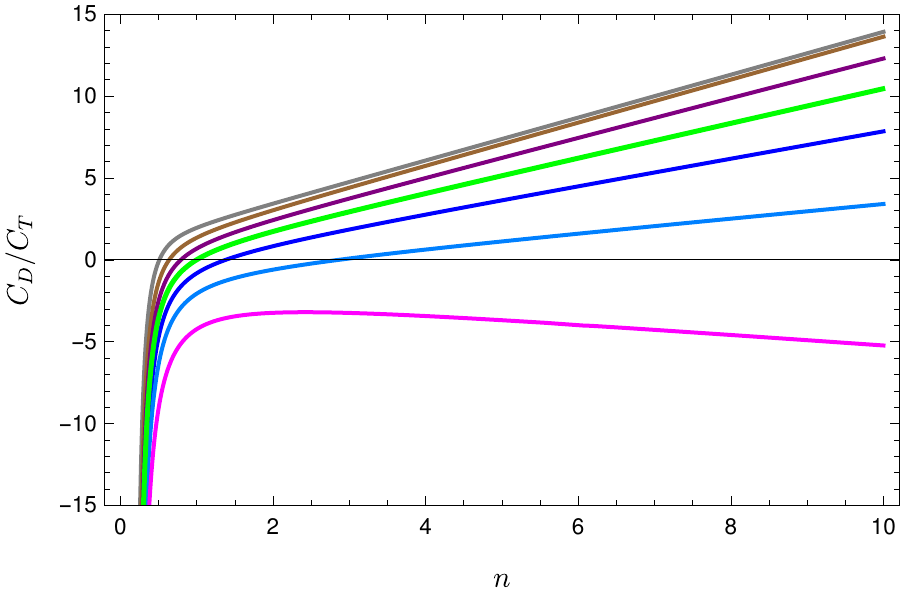}} 
\subfigure[]
{ \includegraphics[scale=0.82]{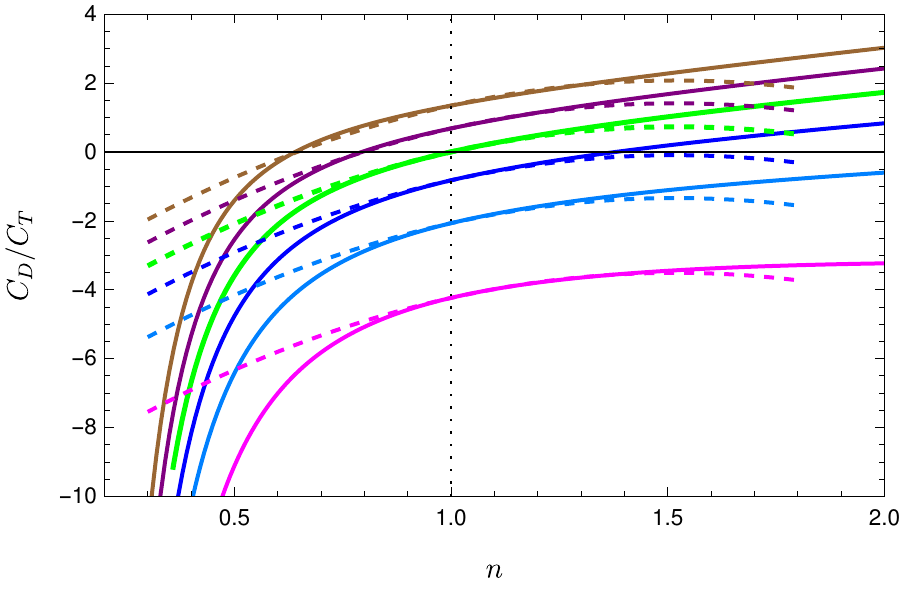}} \\
\caption{Plots of $C_D/C_T$, defined in eqs.~\eqref{eq:CTCV_holography} and \eqref{eq:holographic_equation_CD}, as a function of $n$ in $d=6.$ The colours correspond to different fixed values of the chemical potential $\tilde{\mu}$ in eq.~\eqref{eq:definition_mutilde} with $\ell_{*}/L=1.$ 
(a) The different curves correspond to $\tilde{\mu}^2 =\pm 0.04, \pm 0.09, \pm 0.16$, with $\tilde{\mu}^2$ increasing from top to bottom. We highlight in green the curve corresponding to $\tilde{\mu}=0$. Note that $\tilde{\mu}^2$ can be negative, as the chemical potential can take imaginary values.
 (b) Zoom of the same plot around the region close to $n=1$. The dashed lines correspond to the analytic expansions in eq.~\eqref{eq:analyitic_expansions_CD_over_CT_d6}. Curves with the same colours correspond to the same chemical potential.}
\label{fig:CD6}
\end{figure}

The numerical results in higher dimensions confirm the picture that we outlined in section \ref{sect-numerical} in $d=3,4.$
The plots of $C_D$ as a function of the Rényi index at fixed $\tilde{\mu},$ depicted in fig.~\ref{fig:CD5} and \ref{fig:CD6} for $d=5,6$ respectively, show that an imaginary chemical potential increases the value of $C_D,$ while a real chemical potential decreases it compared to the case without a global symmetry in the system $(\tilde{\mu}=0).$
Furthermore, past a certain value of real $\tilde{\mu},$ the function $C_D$ becomes negative along all the range of the Rényi index.

\begin{figure}[H]
\centering
\subfigure[]
{ \includegraphics[scale=0.82]{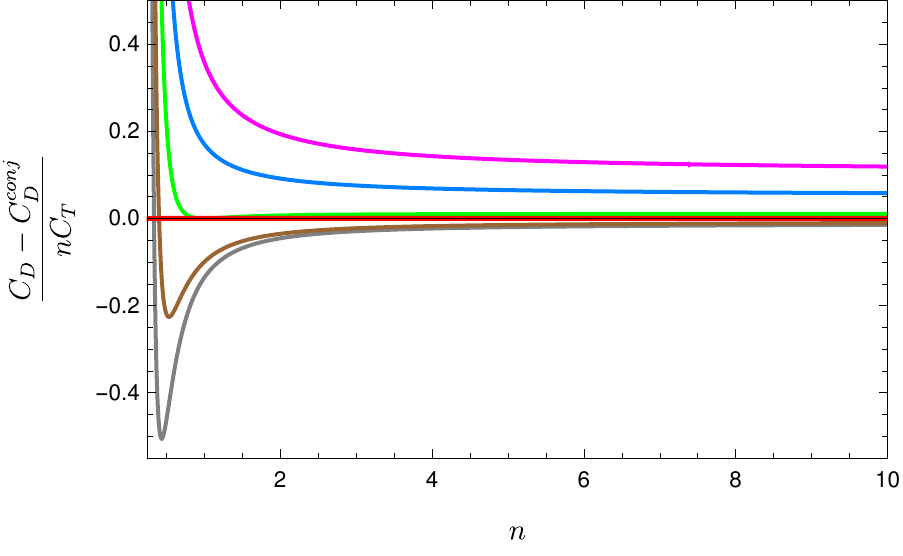}} 
\subfigure[]
{ \includegraphics[scale=0.82]{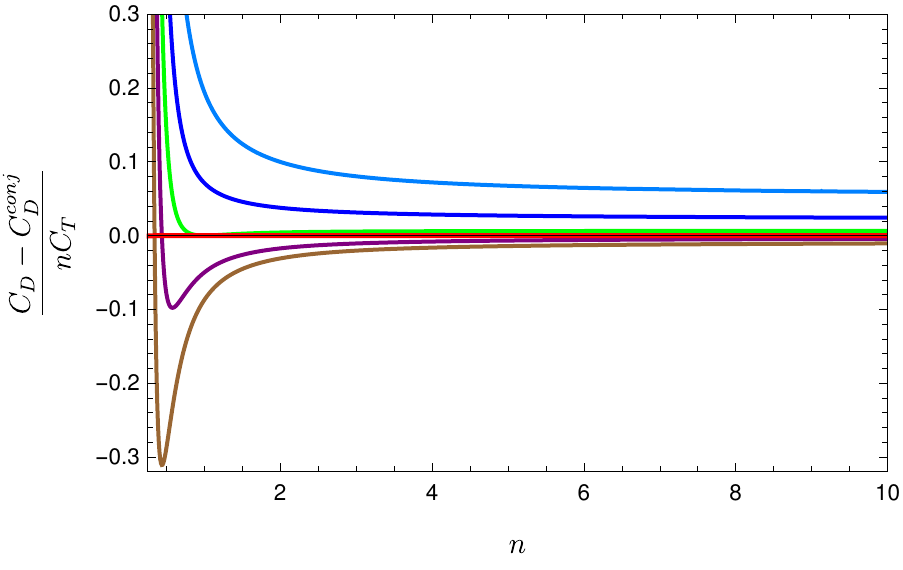}} \\
\caption{Relative difference $(C_D - C_D^{\rm conj})/(n C_T)$ as a function of $n$ for fixed chemical potential in $d=5$ (left) and $d=6$ (right). The different curves correspond to different chemical potentials, $\tilde{\mu}^2 = \pm 0.04, \pm 0.09$, with $\tilde{\mu}^2$ increasing from bottom to top (we set $\ell_{*}/L=1$). We plot the curve with $\tilde{\mu}=0$ in green. Note that $\tilde{\mu}^2$ can be negative, as the chemical potential can take imaginary values.
The red curve corresponds to evaluating the SUSY solution. As it can be seen, in this case $C_D$ is exactly $C_D^{\rm conj}$, at least up to numerical errors of the order $10^{-4}$ in $d = 5,6.$ }
 \label{fig:CD5_rel}
\end{figure}

In fig.~\ref{fig:CD5_rel}  we test explicitly the conjecture \eqref{eq:conj_CD}.
Again, we clearly observe that it is violated at generic values of $n, \tilde{\mu},$ while it is respected in the supersymmetric case, represented by the red curve.
We point out that the numerics becomes more involved for small values of the chemical potential in $d=7.$
Therefore, we avoid to report such plots, but the qualitative behaviour is exactly the same as in the other dimensions.

\bibliographystyle{JHEP}

\bibliography{bibliography}

\end{document}